\documentclass[11pt,letterpaper]{article}
\usepackage[margin=1in]{geometry}

\usepackage{todonotes}
\usepackage{lineno}
\usepackage{tikz}
\usetikzlibrary{shapes.geometric}
\usetikzlibrary{arrows.meta}
\usepackage{caption}

\usepackage[T1]{fontenc}
\usepackage[utf8]{inputenc}
\usepackage{lmodern}
\usepackage{microtype}

\usepackage{amsmath,amssymb,amsthm}
\usepackage{mathtools}
\usepackage{thmtools}
\usepackage{thm-restate}
\usepackage{stmaryrd} 
\usepackage[normalem]{ulem} %

\makeatletter
\@ifundefined{newcounteralias}{}{%
  \renewcommand\thmt@autorefsetup{%
    \@xa\def\csname\thmt@envname autorefname\@xa\endcsname
      \@xa{\thmt@thmname}%
  }%
}
\makeatother

\usepackage{graphicx}
\usepackage[dvipsnames]{xcolor}
\usepackage{booktabs,array,tabularx}
\usepackage{enumitem}
\usepackage{algorithm}
\usepackage{algpseudocode}
\usepackage{tikz}
\usetikzlibrary{arrows.meta,positioning,calc,fit}
\tikzset{
  digraph vertex/.style={circle,draw,inner sep=1.5pt,minimum size=5mm},
  digraph arc/.style={-{Stealth[length=2mm]}}
}

\usepackage{fix-cm}
\usepackage{adjustbox}

\usepackage[hidelinks]{hyperref}
\usepackage[capitalize,nameinlink,noabbrev]{cleveref}

\newcommand{\EulerDeclareTheorem}[2]{%
  \ifcsname #2\endcsname
  \else
    \declaretheorem[#1]{#2}%
  \fi
}
\EulerDeclareTheorem{numberwithin=section,name=Theorem}{theorem}
\EulerDeclareTheorem{sibling=theorem,name=Lemma}{lemma}
\EulerDeclareTheorem{sibling=theorem,name=Proposition}{proposition}
\EulerDeclareTheorem{sibling=theorem,name=Corollary}{corollary}
\EulerDeclareTheorem{sibling=theorem,name=Conjecture}{conjecture}
\EulerDeclareTheorem{sibling=theorem,name=Claim}{claim}
\EulerDeclareTheorem{style=definition,sibling=theorem,name=Definition}{definition}
\EulerDeclareTheorem{style=definition,sibling=theorem,name=Assumption}{assumption}
\EulerDeclareTheorem{style=definition,sibling=theorem,name=Example}{example}
\EulerDeclareTheorem{style=definition,sibling=theorem,name=Counterexample}{counterexample}
\EulerDeclareTheorem{style=remark,sibling=theorem,name=Remark}{remark}
\EulerDeclareTheorem{style=remark,sibling=theorem,name=Observation}{observation}

\crefname{assumption}{Assumption}{Assumptions}
\crefname{counterexample}{Counterexample}{Counterexamples}
\crefname{observation}{Observation}{Observations}
\crefname{algorithm}{Algorithm}{Algorithms}
\crefname{claim}{Claim}{Claims}
\crefname{theorem}{Theorem}{Theorems}
\crefname{lemma}{Lemma}{Lemmas}
\crefname{subsection}{Subsection}{Subsections}

\DeclarePairedDelimiterX{\setof}[2]{\lbrace}{\rbrace}{%
  #1\,\delimsize\vert\,#2}

\DeclareMathOperator{\tw}{tw}

\DeclareMathOperator{\cyc}{cyc}

\newcommand{\cc}{\mathrm{cc}}

\newcommand{\problemname}[1]{\textnormal{\textsc{#1}}}
\newcommand{\strongbackt}{%
    \problemname{Strong Backdoor to Treewidth-$t$}}
\newcommand{\strongback}{%
    \problemname{Strong Backdoor to Forest}}

\newcommand{\gsapprox}{Gaspers and Szeider~\cite{DBLP:conf/focs/GaspersS13}}

\newcommand{\FPT}{\ensuremath{\mathsf{FPT}}}
\newcommand{\Wone}{\ensuremath{\mathsf{W[1]}}}
\newcommand{\Wtwo}{\ensuremath{\mathsf{W[2]}}}

\DeclareMathOperator{\inc}{inc}
\DeclareMathOperator{\var}{var}
\DeclareMathOperator{\core}{core}
\newcommand{\semireduct}[1]{\llbracket #1 \rrbracket}

\newcommand{\defparproblem}[4]{
    \vspace{3mm}
    \noindent\fbox{
        \begin{minipage}{.95\textwidth}
            \begin{tabular*}{\textwidth}{@{\extracolsep{\fill}}lr} \textsc{#1}  & {\bf{Parameter:}} #3 \\ \end{tabular*}
            {\bf{Input:}} #2  \\
            {\bf{Question:}} #4
        \end{minipage}
    }
    \vspace{2mm}
}

\newcommand{\defproblem}[3]{
    \vspace{3mm}
    \noindent\fbox{
        \begin{minipage}{.95\textwidth}
            \begin{tabular*}{\textwidth}{@{\extracolsep{\fill}}lr} \textsc{#1}   \\ \end{tabular*}
            {\bf{Input:}} #2  \\
            {\bf{Question:}} #3
        \end{minipage}
    }
    \vspace{2mm}
}

\newcommand{\cla}{\mathrm{cla}}

\title{Strong Backdoor to Forests is FPT:\\
Tight Linear-Time FPT for Essentially General SAT}

\author{
  Olesia Bilyk\textsuperscript{1}
  \quad
  Aidar Munduzbaev\textsuperscript{2}
  \quad
  Roohani Sharma\textsuperscript{3}\thanks{Supported by the Young Scientist Fellowship (YSF) of the Institute for Basic Science (IBS-R029-Y8).}
  \quad
  Mamnoon Siam\textsuperscript{2}
  \\[1.2em]
  \normalsize
  \textsuperscript{1}Independent Researcher\\
  \normalsize
  \textsuperscript{2}Korea Advanced Institute of Science \& Technology (KAIST),
  Daejeon, South Korea\\
  \normalsize
  \textsuperscript{3}DIMAG, Institute for Basic Science,
  Daejeon, South Korea\\[0.6em]
  \small
  \texttt{olesia.o.bilyk@gmail.com, \{amunduzbaev,mamnoon\}@kaist.ac.kr, roohani@ibs.re.kr}
}
\date{}

\begin{document}

\maketitle
\thispagestyle{empty}

\vspace{-8mm}

\begin{abstract}
The Boolean Satisfiability (SAT) problem is NP-complete, yet practical SAT solvers work exceptionally well in practice.
Williams, Gomes, and Selman [IJCAI 2003]
introduced \emph{backdoors} to help explain this gap.
Backdoors are small sets of variables whose assignment results in a formula in some fixed tractable class.
We study the problem of finding strong backdoors to formulas whose incidence
graphs have treewidth at most a prescribed bound $t$: find a small set of variables such that every assignment to this set of variables results in a formula whose incidence graph has treewidth bounded by $t$.

In a seminal work, Gaspers and Szeider
[FOCS 2013] gave a fixed-parameter tractable (FPT) approximate algorithm
for finding a strong backdoor of size at most $k$ to incidence treewidth at most $t$, for a given
CNF-formula $\phi$.
In particular, their algorithm runs 
in $f(k,t) \cdot |\phi|^3$ time, and finds a strong backdoor to treewidth at most $t$,
of $\phi$, of size
at most $2^k$ whenever there exists one of size at most $k$.
Determining whether the detection of strong backdoors to bounded incidence treewidth is FPT parameterized by $k+t$ was left open in their work and this has been repeatedly posed as an open question by Gaspers and Szeider, for example in [The Multivariate Algorithmic Revolution and Beyond 2012] and more recently in [Computer Science Reviews 2025],
even for incidence forests and when the input is a $3$-CNF formula.
We answer this question affirmatively.

There are several interesting features of our result, beyond resolving the open question of Gaspers and Szeider.
First, Lokshtanov, Panolan, and Ramanujan [ICALP 2022]
proved \Wtwo-hardness of finding strong backdoors even when the target class is the class of forests, for arbitrary CNF formulas. 
We observe that their hardness construction crucially relies on 
using tautological clauses in the formula, that is clauses containing a variable and its negation,
and does not establish hardness when such clauses are excluded.
For the purposes of solving SAT, such tautological clauses in the input instance are meaningless.

We show that strong backdoors to forests can be found in linear-FPT time
on \emph{arbitrary non-tautological} CNF formulas (not just on $3$-CNF formulas).
Given such a formula
$\phi$ and a budget $k$, our algorithm finds a strong backdoor to the class of forests,
of size at most $k$, or correctly reports that none exists,
in $2^{2^{O(k)}} \cdot |\phi|$ time. 
Moreover, we show that our result is as tight as it can get.
First, we show that finding strong backdoors to treewidth $t$ for any $t \geq 2$ is \Wone-hard, even on non-tautological $3$-CNF formulas.
Second, the running time of our algorithm is tight: it is linear in the input size and a better than double-exponential dependence on $k$ in the running time would imply a $2^{o(N/\log N)}$-time algorithm for solving {\sc $3$-SAT} on $N$ variables, which would beat even the very recently claimed sub-exponential algorithm for solving {\sc $3$-SAT} by OpenAI and Anthropic.

\end{abstract}

\newpage

\setcounter{page}{1}
\hypersetup{pageanchor=true}

\section{Introduction}\label{sec:intro}

The Boolean Satisfiability (SAT) problem is a cornerstone of
computational complexity and a central tool for solving search and
reasoning problems. Although SAT is NP-complete, modern SAT solvers
routinely handle large instances arising in applications. Understanding
which properties of these instances make this possible has motivated
extensive research into determining tractable fragments of SAT and ways of exposing
them inside more general formulas. Williams, Gomes, and
Selman~\cite{DBLP:conf/ijcai/WilliamsGS03} introduced \emph{backdoors}
as one explanation for this phenomenon: assigning a small set of key
variables can reduce a large formula to an instance of a tractable
class. A small backdoor can therefore reveal tractability that is not
apparent from the formula as a whole.

The choice of the tractable target class gives rise to two broad
directions, both of which are very well
explored~\cite{DBLP:conf/sat/NishimuraRS04,
DBLP:journals/jcss/GaspersMOSZ17,
DBLP:conf/icalp/GaspersS12,
DBLP:conf/focs/GaspersS13,
DBLP:conf/soda/FominLMRS15,
DBLP:conf/icalp/LokshtanovP022} (see related work paragraph).
\emph{Syntactic} classes of formulas restrict the structure of the clauses.
Standard examples include Horn formulas, in which each clause has at
most one positive literal, and $2$-CNF formulas, in which each clause
has at most two literals. 
On the other hand \emph{structural} classes restrict
the interactions between variables and clauses, typically through a
graph representation of the formula. 
A fundamental example is the notion of (bounded)
\emph{incidence treewidth} of a formula. 
The incidence graph of a CNF formula has
one vertex for each variable and each clause, with an edge joining a
variable to every clause in which it occurs (positively or negatively). 
For every fixed $t$, SAT
is solvable in linear time on formulas whose incidence graphs have
treewidth at most $t$~\cite{DBLP:journals/jda/SamerS10}.

There are two standard notions of backdoors to a target class
$\mathcal{C}$. A set $X$ of variables is a \emph{weak backdoor} if
some assignment $\tau$ to $X$ leaves a satisfiable formula
$\phi[\tau]\in\mathcal{C}$. 
It is a \emph{strong backdoor} if
$\phi[\tau]\in\mathcal{C}$ for every assignment $\tau$ to $X$.
Here, $\phi[\tau]$ is obtained by deleting satisfied clauses and
removing falsified literals, with identical resulting clauses merged.
Given a strong backdoor of size $k$ for a formula $\phi$, 
one can decide satisfiability of $\phi$ by
solving the $2^k$ resulting formulas obtained by fixing an assignment to the $k$ strong backdoor variables, each of which belongs to a structured target class of choice. Consequently,
a small strong backdoor to bounded incidence treewidth permits
efficient SAT solving even when the original incidence graph has
unbounded treewidth.

To exploit this approach towards solving {\sc SAT}, 
however, one must first find such a backdoor.
This leads to the central algorithmic question of \emph{(strong/weak) backdoor
detection}: given a formula $\phi$ and an integer $k$, can we find a 
(strong/weak) backdoor of size at most $k$, or correctly determine that none exists,
in time $f(k) \cdot |\phi|^{O(1)}$ for a computable function $f$? 
Such an
algorithm is \emph{fixed-parameter tractable} (FPT) with respect to
$k$.
For bounded-treewidth target class, one may also include the target
width $t$ in the parameter and ask for an algorithm with running time $f(k,t) \cdot |\phi|^{O(1)}$. Backdoor detection and its parameterized
complexity have become a substantial research program, see for example the surveys
of Gaspers and
Szeider~\cite{DBLP:conf/birthday/GaspersS12,DBLP:journals/csr/GaspersS26}, one of which is as recent as from last year and the other one is from more than a decade ago.

In this work we focus on finding strong backdoors to bounded
incidence treewidth. 
Gaspers and
Szeider~\cite{DBLP:conf/icalp/GaspersS12} first obtained an
FPT-approximation algorithm for strong backdoors to the class of forests (treewidth $1$). They subsequently, in a seminal work,
extended this result to the case when the target class has arbitrary
bounded incidence treewidth~\cite{DBLP:conf/focs/GaspersS13}.
Their result showed that
given a CNF formula $\phi$
and integers $k,t$, there is an  algorithm that runs in $f(k,t) \cdot |\phi|^3$ time, for some computable function $f$,
that either finds a strong backdoor to incidence treewidth at most $t$
of size at most $2^k$, or correctly reports that no such backdoor of
size at most $k$ exists. 
Such an FPT-approximation algorithm suffices to establish fixed-parameter
tractability of SAT parameterized by the size of a smallest strong
backdoor to the target class of incidence treewidth at most $t$. 

The work of Gaspers and Szeider~\cite{DBLP:conf/focs/GaspersS13} leaves open a fundamental research question:

\begin{center}
{\em Is finding strong backdoors to treewidth-$t$ FPT parameterized by $k+t$?}
\end{center}
That is, is there an algorithm that runs in $f(k,t) \cdot |\phi|^{O(1)}$ time and either reports a strong backdoor to treewidth at most $t$ of size at most $k$, or correctly concludes that none exists?

\paragraph{Weak backdoors.}
For weak backdoor detection, even when the target class is the class of forests (that is when the incidence treewidth is $1$) the problem is hard. 
In particular,
Gaspers and Szeider~\cite{DBLP:conf/icalp/GaspersS12} proved that
finding a weak backdoor to forests is \Wtwo-hard (on unrestricted CNF formulas), parameterized
by $k$. 
In fact, unlike the strong backdoors case, unless $\FPT=\Wone$, the problem admits no FPT-approximation whose approximation factor is bounded by any computable function
of $k$,
even when the target class is that of forests.\footnote{The reduction of Gaspers and
Szeider~\cite{DBLP:conf/icalp/GaspersS12} from
\textsc{Hitting Set} preserves the optimum solution size.
Composing it with the standard optimum-preserving reduction
from \textsc{Dominating Set} to \textsc{Hitting Set}, and
applying the FPT-inapproximability result of Karthik,
Laekhanukit, and Manurangsi~\cite{DBLP:journals/jacm/SLM19},
yields the claim.}
Positive FPT results for the question of finding weak backdoors to bounded incidence treewidth classes, have thereafter focused on restrictions on the structure of the
input formula. 
Fomin, Lokshtanov, Misra, Ramanujan, and
Saurabh~\cite{DBLP:conf/soda/FominLMRS15} obtained a linear-time FPT
algorithm for solving $d$-SAT via backdoors to bounded treewidth,
and established FPT-time detection of weak backdoors when the input formula is a $d$-SAT formula for a fixed constant $d$.
Later, Lokshtanov, Panolan, and
Ramanujan~\cite{DBLP:conf/icalp/LokshtanovP022} extended the FPT algorithm for weak-backdoor
detection to bounded incidence treewidth class, when the input formula's incidence graph exclude a fixed biclique $K_{d,d}$
as a subgraph. 
Their result applies, in particular, when the incidence graph of the input formula belongs to the class of nowhere-dense graphs.
More recently, Dreier, Ganian, Gr\"undel,
Sharma and Wietheger~\cite{dreier2026solvingsattreewidthbackdoors} obtained
a linear-time FPT algorithm for finding a weak backdoor to incidence treewidth $t$, when the signed incidence graph of the input formula is
semi-ladder-free. This extends tractability to classes of input formulas whose signed incidence graphs belongs to an appropriate dense graph class. 
The results mentioned in this paragraph
concern finding weak backdoors in FPT time.
The result of Fomin, Lokshtanov, Misra, Ramanujan, and
Saurabh~\cite{DBLP:conf/soda/FominLMRS15} also deals with strong backdoors but essentially gives an FPT-approximation like Gaspers and Szeider~\cite{DBLP:conf/focs/GaspersS13} for strong backdoors, but for $d$-SAT formulas in {\em linear} FPT time.

\paragraph{Strong backdoors.}
Coming back to the topic of finding strong backdoors to incidence treewidth $t$, which is the topic of this work, unlike for finding weak backdoors, 
the existence of an exact FPT algorithm has remained open
, even for $3$-CNF formulas, and even when the target class is the class of incidence treewidth $1$ (that is forests).
Asked more than a decade ago~\cite{DBLP:conf/focs/GaspersS13,DBLP:conf/birthday/GaspersS12}, this specific question
was reiterated in the recent survey of Gaspers and
Szeider~\cite{DBLP:conf/birthday/GaspersS12,DBLP:journals/csr/GaspersS26}, which we resolve in positive in this work.

\begin{center}
{\em Is finding a strong backdoor to the class of forests, given a $3$-CNF formula, FPT?} 
\end{center}

Lokshtanov, Panolan and Ramanujan~\cite{DBLP:conf/icalp/LokshtanovP022} proved \Wtwo-hardness of
finding strong backdoors to incidence treewidth $t$, for every fixed $t \geq 1$ (that is, including the case when the target class is that of forests),
on {\em arbitrary} CNF formulas. There is, however, a crucial feature of
their construction that is worth making a note of.
The constructed hard instance relies on having \emph{tautological clauses}, that is clauses that contain
both a variable and its negation and therefore are satisfied under every assignment. 
Though such tautological clauses can non-trivially affect the problem of finding (strong) backdoors,
for the purposes of understanding SAT, which is the eventual goal of studying backdoors at the first place, 
such clauses are useless as almost any pre-processing algorithm for SAT will first get rid of these clauses before doing anything non-trivial on the instance.
Thus, despite the hardness result of Lokshtanov, Panolan, and Ramanujan~\cite{DBLP:conf/icalp/LokshtanovP022}, the following question still remained widely open.

\begin{center}
{\em Is finding a strong backdoor to the class of incidence treewidth $t$, given a non-tautological CNF formula, FPT?} 
\end{center}

\paragraph{Our results.}
We resolve this question positively  for $t =1$ and negatively for any $t \geq 2$.
More formally,
for every fixed integer $t\ge1$, we consider the following problem.

\defparproblem
    {Strong Backdoor to Treewidth-$t$}
    {A non-tautological CNF formula $\phi$ and an integer $k\ge0$.}
    {$k$}
    {Does $\phi$ have a strong backdoor to treewidth-$t$
     of size at most $k$?}

For $t=1$, we call this problem \strongback.

\begin{restatable}{theorem}{strongforestfpt}
\label{thm:main}
\strongback{} can be solved in $2^{2^{O(k)}} \cdot |\phi|$ time.
\end{restatable}

\cref{thm:main}, in particular resolves the open question posed by Gaspers and Szeider~\cite{DBLP:conf/birthday/GaspersS12,DBLP:journals/csr/GaspersS26} about finding strong backdoors to forests on $3$-CNF formulas, and goes much beyond its resolution.
When the input formula is a $3$-CNF formula, 
the non-tautological assumption
can be dropped; see
Appendix~\ref{sec:tautological-3cnf} after reading Sections~\ref{sec:preliminaries} and~\ref{sec:algo}.
Thus our result also answers the question for
$3$-SAT~\cite{DBLP:conf/birthday/GaspersS12,DBLP:journals/csr/GaspersS26} without excluding tautological clauses.

Interestingly, we show that instead the hard obstacle lies in generalizing the treewidth of the target class (and not really in generalizing the structure of the input instance).

\begin{restatable}{theorem}{strongtreewidthhardness}
\label{thm:hardness}
For every fixed integer $t\geq 2$, \strongbackt{} is \Wone-hard
parameterized by $k$, even on non-tautological $3$-CNF formulas.
\end{restatable}
Together with the FPT-approximation of Gaspers
and Szeider~\cite{DBLP:conf/focs/GaspersS13}, \cref{thm:hardness} separates the exact detection of strong backdoors to incidence treewidth $t = 1$, 
from FPT-approximation for
every $t$.

To prove \cref{thm:hardness} we first show that the graph problem {\sc Simultaneous Treewidth-$t$ Deletion} is \Wone-hard whenever $t \geq 2$.
Interestingly, 
this problem was shown to be FPT when $t=1$ in~\cite{DBLP:journals/toct/AgrawalLMS18}. 

\defparproblem{\textsc{Simultaneous Treewidth-$t$ Deletion}}
  {Graphs $G_1,\ldots,G_{\alpha}$ on the same vertex set $V$,
   and a nonnegative integer $K$.}
  {$K+\alpha$}
  {Does there exist $D\subseteq V$ with $|D|\le K$ such that
   $\operatorname{tw}(G_a-D)\le t$ for every $a\in[\alpha]$?}

\begin{restatable}[Hardness of {\sc Simultaneous Treewidth-$t$ Deletion}]{lemma}{lemmasimtw}\label{lem:simultaneous-treewidth-hardness}

For every fixed integer $t\ge2$,
\textsc{Simultaneous Treewidth-$t$ Deletion} is \Wone-hard
parameterized by $K+\alpha$.

\end{restatable}

Finally, we first observe that the running time of \cref{thm:main} is linear in the size of the input, which is already optimum, and at same time we show that the double-exponential dependence on $k$ in the running time is ``unavoidable'', as a better than double-exponential dependence on $k$ for \strongback\
would solve {\sc $3$-SAT} on $n$ variables in $2^{o(n/\log n)}$ time.
This would have refuted the Exponential Time Hypothesis (ETH).
But already a few days ago,
OpenAI~\cite{OpenAI2026ETH} and Anthropic~\cite{Claude2026ETH}
announced that the Exponential Time Hypothesis (ETH) is false, by giving an algorithm that solve {\sc $3$-SAT} in $2^{O(n \log \log n / \sqrt{\log n})}$ time.
Our ``lower bound'' now implies that 
a better than double-exponential dependence in $k$ in the algorithm of \cref{thm:main}
would give an even faster algorithm for {\sc $3$-SAT} than what is currently claimed by OpenAI and Anthropic.

\begin{restatable}{theorem}{strongforesteth}
\label{thm:eth}
Unless {\sc $3$-SAT} on $n$ variables can be solved in $2^{o(n/\log n)}$ time,
\strongback{} cannot be solved in
$2^{2^{o(k)}} \cdot | \phi|^{O(1)}$ time, even on non-tautological
$3$-CNF formulas.
\end{restatable}

\paragraph{Related work.}
Finding backdoors in FPT time has been an active research direction
for over two decades, with continuing work on tractable target
classes and the complexity of detecting backdoors into them.
The surveys of Gaspers and
Szeider~\cite{DBLP:conf/birthday/GaspersS12,DBLP:journals/csr/GaspersS26}
give comprehensive accounts of this development and the parameterized
complexity of backdoor detection.

Beyond the bounded-treewidth target classes discussed in this work, syntactic
targets have received extensive attention. 
Nishimura, Ragde, and
Szeider~\cite{DBLP:conf/sat/NishimuraRS04} proved that strong-backdoor detection
into Horn and $2$-CNF is FPT, whereas weak detection is \Wtwo-hard.
Misra, Ordyniak, Raman, and Szeider~\cite{DBLP:conf/sat/MisraORS13}
obtained single-exponential
algorithms for weak detection on $3$-CNF formulas.
Gaspers, Ordyniak, Ramanujan, Saurabh, and
Szeider~\cite{DBLP:journals/algorithmica/GaspersORSS16}
proved that detecting weak or strong backdoors to $q$-Horn
is \Wtwo-hard and gave an FPT-approximation algorithm
for finding deletion backdoors to $q$-Horn.
Ramanujan and Saurabh~\cite{DBLP:conf/soda/RamanujanS14}
gave an exact FPT algorithm for deletion-backdoor detection\footnote{A deletion-backdoor is one where deleting all literals of this set from all clauses results in the mentioned tractable class.}
with
linear dependence on the input size.
Gaspers, Misra, Ordyniak, Szeider, and
\v{Z}ivn\'y~\cite{DBLP:journals/jcss/GaspersMOSZ17} classified strong backdoor detection to classes of formulas that are
 unions of Schaefer classes. For example, Horn $\cup\,2$-CNF
admits FPT detection, whereas Horn $\cup$ dual Horn is \Wtwo-hard.

Gaspers and Szeider~\cite{DBLP:conf/sat/GaspersS12} obtained an FPT-approximation for strong backdoor detection into nested formulas.
Ordyniak, Paulusma, and Szeider~\cite{DBLP:journals/tcs/OrdyniakPS13} proved
polynomial-time solvability of SAT on $\beta$-acyclic formulas,
but \Wtwo-hardness of strong-backdoor detection into this class.
Here acyclicity is defined on the formula's hypergraph, giving a
target class more general than incidence forests.

Related work also studies parameters that can remain bounded when
minimum backdoor size is unbounded.
Ganian, Ramanujan, and Szeider~\cite{DBLP:conf/sat/GanianRS17,DBLP:conf/stacs/GanianRS17}
obtained FPT detection and evaluation algorithms using the torso
treewidth of a backdoor into suitable SAT and CSP targets.
M\"ahlmann, Siebertz, and Vigny~\cite{DBLP:conf/mfcs/MahlmannSV21} introduced
recursive backdoor depth, which permits independent processing of
components created by assignments and parallels recursive
elimination in treedepth. Dreier, Ordyniak, and
Szeider~\cite{DBLP:journals/jcss/DreierOS24,DBLP:journals/constraints/DreierOS23} developed this approach
for syntactic SAT targets and CSP, including FPT-approximation of
depth and linear-time SAT solving for bounded depth into Horn,
dual Horn, and $2$-CNF.
Adaptive backdoor trees~\cite{DBLP:conf/aaai/SamerS08} and backdoor
DNFs~\cite{DBLP:journals/jcss/OrdyniakSS24} provide further detection problems
parameterized by the number of leaves or terms.
These directions vary the parameter or certificate; our work
concerns exact detection of small backdoors into a structural
target class.

\section{Preliminaries}\label{sec:preliminaries}
For a nonnegative integer $q$, write
$[q]:=\{1,\ldots,q\}$.
All logarithms are to base two.
Graphs are finite, undirected, and simple unless
stated otherwise.

\paragraph{Graph notation.}
For a graph $G$, write $V(G)$ and $E(G)$ for its
vertex and edge sets, respectively.
We write $\deg_G(v)$ for the degree of a vertex $v$
and $\tw(G)$ for the treewidth of $G$.
A graph has treewidth at most one if and only if
it is a forest.

For an integer $d\geq0$, a graph $G$ is
\emph{$d$-degenerate} if every nonempty subgraph
of $G$ contains a vertex of degree at most $d$.
The \emph{degeneracy} of $G$ is the smallest such
integer $d$.

The \emph{$2$-core} of $G$, denoted by $\core_2(G)$,
is the unique maximal induced subgraph in which
every vertex has degree at least two.
It can be computed in linear time by repeatedly
deleting vertices of degree at most one.
The resulting graph is independent of the deletion
order and may be empty.
Every cycle of $G$ lies in $\core_2(G)$.

\begin{definition}[Cyclomatic number]
\label{def:cyclomatic-number}
Let $\cc(G)$ denote the number of connected
components of a graph $G$.
The \emph{cyclomatic number} of $G$ is
\[
    \cyc(G):=
    |E(G)|-|V(G)|+\cc(G).
\]
\end{definition}

Equivalently, $\cyc(G)$ is the minimum number of
edges whose deletion makes $G$ a forest.
In particular, $\cyc(G)=0$ if and only if $G$ is
a forest.
Deleting a vertex of degree $a$ decreases the
cyclomatic number by at most $\max\{a-1,0\}$.
Deleting a vertex of degree at most one leaves
the cyclomatic number unchanged; consequently,
\[
    \cyc(G)=\cyc(\core_2(G)).
\]

\begin{lemma}\label{lem:bounded-number-of-cycles}
Let $c\geq0$ be an integer and let $G$ be a graph
with $\cyc(G)\leq c$.
Then $\core_2(G)$ contains at most $2^c-1$ distinct
simple cycles.
\end{lemma}

\begin{proof}
Let $F$ be a spanning forest of $G$ and put
$R:=E(G)\setminus E(F)$.
Then $|R|=\cyc(G)\leq c$.
Every cycle contains an edge of $R$.
Distinct simple cycles have distinct intersections
with $R$: otherwise, their symmetric difference
would be an even-degree subgraph of the forest $F$
and would therefore be empty.
Thus the cycles of $G$ inject into the nonempty
subsets of $R$, giving at most $2^c-1$ cycles.
The same bound holds for $\core_2(G)$.
\end{proof}

\paragraph{CNF formulas and incidence graphs.}
A \emph{literal} is a Boolean variable $x$ or its
negation $\neg x$.
These are called positive and negative literals,
respectively.

A \emph{clause} is a finite set of literals,
interpreted as their disjunction.
A \emph{CNF formula} is a finite set of distinct
clauses, interpreted as their conjunction.
The empty clause is false, and the empty formula
is true.

A clause is \emph{tautological} if it contains both
$x$ and $\neg x$ for some variable $x$.
A formula is \emph{non-tautological} if none of its
clauses is tautological.

For a clause $C$, let $\var(C)$ denote its set of
variables.
For a formula $\phi$, put
\[
    \var(\phi):=\bigcup_{C\in\phi}\var(C).
\]
We may regard $\phi$ as a formula over any 
variable set $V\supseteq\var(\phi)$; variables of $V$
need not occur in $\phi$.

The \emph{incidence graph} $\inc(\phi)$ is the
bipartite graph whose two parts consist of the
variable vertices $\var(\phi)$ and the clause
vertices of $\phi$.
A variable vertex $x$ is adjacent to a clause
vertex $C$ if and only if $x\in\var(C)$.
We denote the set of clause vertices by $\cla(\phi)$
and identify each clause vertex with its clause
when convenient.
The incidence edge $xC$, where $x$ is a variable vertex and $C$ is a clause vertex,
is labelled positive if $x\in C$
and negative if $\neg x\in C$.
For a non-tautological formula, every incidence
has exactly one sign.

We measure the size of a formula by
\[
    |\phi|:=
    |\var(\phi)|+|\cla(\phi)|
    +\sum_{C\in\phi}|C|,
\]
where $|C|$ is the number of literals in $C$.

\paragraph{Assignments and reducts.}
Let $\phi$ be a formula over a variable
set $V$.
An \emph{assignment} to $X\subseteq V$ is a function
$\tau:X\to\{0,1\}$.
For $x\in X$, the literal $x$ is true under $\tau$
if $\tau(x)=1$, and $\neg x$ is true if $\tau(x)=0$.
A clause is satisfied by $\tau$ if it contains
a literal that is true under $\tau$.
Assignments to variables absent from $\phi$
have no effect on $\phi$.

The \emph{semi-reduct} $\phi\semireduct{\tau}$ is
obtained by deleting every clause satisfied by
$\tau$ and removing all literals on variables
of $X$ from the remaining clauses, without
merging clauses that become identical.
Each retained clause keeps its original identity.
In particular, clauses that become empty are retained.

The graph $\inc(\phi\semireduct{\tau})$ has one
clause vertex for each retained clause and one
variable vertex for each variable in
$\var(\phi)\setminus X$, including isolated
variable vertices.
Thus $\inc(\phi\semireduct{\tau})$ is a subgraph
of $\inc(\phi)$.
An empty clause corresponds to an isolated
clause vertex.

The \emph{reduct} $\phi[\tau]$ is the CNF formula
obtained from $\phi\semireduct{\tau}$ by merging
identical clauses.
When $X=\{v\}$ and $\tau(v)=b$, we also write
$\phi[v=b]$.
For consistent assignments $\tau$ and $\rho$,
$\phi[\tau][\rho]=\phi[\tau\cup\rho]$.

\paragraph{Strong backdoors.}
Let $\phi$ be a CNF formula over $V$, and let
$t\geq1$ be an integer.
A set $X\subseteq V$ is a
\emph{strong backdoor to treewidth-$t$} of $\phi$
if
$\tw(\inc(\phi[\tau]))\leq t$
for every assignment $\tau:X\to\{0,1\}$.
When $t=1$, we call $X$ a \emph{strong forest backdoor}.

\paragraph{Convention on reducts.}
Two distinct clauses of a CNF-formula $\phi$ can become identical after a partial
assignment has been fixed. The semi-reduct $\phi\semireduct{\tau}$ keeps
a separate copy for each surviving original clause, and hence would keep multiple copies for clauses that become identical after fixing the assignment $\tau$.
Since we represent a CNF formula as a set of clauses, we
form the reduct $\phi[\tau]$ by merging identical copies.

The two representations (reducts and semi-reducts) describe the same Boolean function,
but they can give different incidence graphs. For example,
setting $x=y=0$ in $(a\lor b\lor x)\land(a\lor b\lor y)$
leaves two copies of $a\lor b$ in the semi-reduct.
The two clause vertices, together with $a$ and $b$, form
a four-cycle. In the reduct, the copies merge into a
single clause vertex, and the incidence graph is a path.

Gaspers and Szeider formulate the open question of exact
strong forest backdoor detection using reducts in both
their original survey and its follow-up
~\cite{DBLP:conf/birthday/GaspersS12,DBLP:journals/csr/GaspersS26}.
That is, they ask whether there is a set of variables $X$ such that for every assignment $\tau$ to $X$, the incidence graph of the  resulting {\em reduct} $\phi[\tau]$ has treewidth at most $t$. This is the version that we address in this paper.
Though the formulation underlying their FOCS 2013 FPT-approximation
algorithm~\cite{DBLP:conf/focs/GaspersS13} uses {\em semi-reducts}: they require the treewidth of the incidence graph of $\phi\semireduct{\tau}$ to have treewidth at most $t$.
We note that working with reducts instead of semi-reducts, only adds more complications to be handled in the proof:
a cycle can disappear even if the assignment fixes none of
its variables and satisfies none of its clauses, in the reduct notation. 
We remark that our proof can be traced to also give the same result if working with semi-reducts instead of reducts.

\paragraph{Representation and running times.}
Running times are measured in the word-RAM model.
Variables, literals, and graph vertices are represented
by integer indices.
Graphs and formulas are stored using adjacency
and literal-occurrence lists.

\paragraph{Integer sorting.}
We use {\em counting sort} to sort $N$ records whose
integer keys belong to $\{0,\ldots,n\}$, in
$O(N+n)$ time~\cite{DBLP:books/mg/CormenLRS01}.
Counting sort preserves the relative input order
of records with the same key.

For records with two keys $(a,b)$, both in
$\{0,\ldots,n\}$, first sort by $b$ and then by $a$
using counting sort.
The second sorting step preserves the ordering
by $b$ among records with the same $a$.
Thus records with identical pairs of keys occur
consecutively, and the total time is $O(N+n)$.
More generally, records with $h$ keys, each in
$\{0,\ldots,n\}$, can be sorted by processing
the keys from last to first, in $O(h(N+n))$ time.

\section{Technical overview}\label{sec:tech-overview}

We begin with a technical overview of the proof of
\cref{thm:main}, which gives an algorithm for finding
a strong forest backdoor of size at most $k$ in a
non-tautological CNF formula $\phi$ in
$2^{2^{O(k)}}\cdot|\phi|$ time.
We first recall how the FPT-approximation result of
Gaspers and Szeider~\cite{DBLP:conf/focs/GaspersS13}
is obtained.
Given $\phi$, $k$, and a target treewidth $t$,
their algorithm runs in $f(k,t)\cdot|\phi|^3$ time
and either returns a strong backdoor to incidence
treewidth at most $t$ of size at most $2^k$,
or correctly reports that none of size at most $k$
exists.

\paragraph{The need to solve a ``simultaneous'' version.}
To capture what makes dealing with strong backdoors tricky,
suppose that an algorithm somehow finds a variable $x$ that is guaranteed to be in the sought strong backdoor solution.
How can one get rid of such a variable?
For many problems having an element of the solution at hand lets us delete this element from the instance, put it in the solution, decrease the solution budget $k$ by $1$ and recurse on the resulting instance with a smaller solution budget.
For the strong backdoor detection problem, this almost standard step does not work.
Indeed, if $x$ is in the solution, to get rid of it, we need to remember that in the resulting instance there is a set of $k-1$
strong backdoor variables
in the formula $\phi[x=0]$ (obtained by setting $x=0$ in $\phi$)
and in the formula $\phi[x=1]$ (obtained by setting $x=1$ in $\phi$), {\em simultaneously}.

The reason the algorithm of Gaspers and Szeider~\cite{DBLP:conf/focs/GaspersS13} introduces approximation, is that they deal with the above scenario by looking at the two formulas $\phi[x=0]$ and $\phi[x=1]$ {\em independently}.
They recursively solve the problem with new parameter $k-1$ now in instance $\phi[x=0]$ and independently in $\phi[x=1]$, and return the union of these two solutions along with $x$.

To avoid getting the approximation, our strategy is to solve the ``simultaneous'' version of the problem, where one is given a set of $\alpha$ formulas $\phi_1, \ldots, \phi_{\alpha}$ on the same variable set, where $\alpha$ at any point is bounded by $2^{O(k)}$,
and the goal is to find a set of at most $k$ variables that is {\em simultaneously} a strong backdoor for each formula $\phi_i$, $i \in [\alpha]$.

\paragraph{The treewidth-based win/win strategy of Gaspers and Szeider.}
To get a variable $x$ that is guaranteed to be in a sought solution,
\gsapprox\ use the machinery from Graph Minors~\cite{RobertsonS1986}. 
The approach involving graph minors is a usual treewidth-based win/win approach: if the treewidth of the (incidence graph) of the formula is large, then they find a bounded set of variables that is guaranteed to intersect every solution, in which case one can branch on this set to get the desired variable $x$.
Otherwise, the treewidth of the formula is small, in which case the application of Courcelle's theorem~\cite{ArnborgLS1991}
solves the problem completely in one shot.

Since we settled to solve the simultaneous version of the problem, to avoid the approximation factor, 
the treewidth-based win/win approach of \gsapprox\ stalls for us.
Recall we now have an instance $(\phi_1, \ldots, \phi_{\alpha},k)$.
Even if one of the formulas $\phi_i$ have large treewidth, we can use the graph minors approach of \gsapprox\ 
to find a small set of variables that is supposed to intersect {\em every} solution of $\phi_i$ and therefore in particular, every solution of the simultaneous instance $(\phi_1, \ldots, \phi_{\alpha},k)$.
Interestingly, the problem lies in the other direction: when the treewidth of each individual formula $\phi_i$ is bounded, one {\em cannot} rely on Courcelle's theorem. And that is not because the simultaneous version is not expressible in CMSO-logic, but that's because even though the treewidth of each formula is bounded, the treewidth of the entire instance $(\phi_1, \ldots, \phi_{\alpha},k)$ containing some $\alpha \geq 2$ formulas may not be bounded.
This is easy to observe: let $G_1$ and $G_2$ be two graphs on the same vertex set where each of them is a disjoint union of several paths. Then each $G_i$ has treewidth $1$ but the union of $G_1$ and $G_2$ can have unbounded treewidth as the union can be a grid if these disjoint paths are chosen appropriately: the set of paths in $G_1$ going from ``left-to-right'' and the ones in $G_2$ going from ``top-to-bottom''.

In this work
we completely avoid the use of the Graph Minors machinery.
As our lower bound of Theorem~\ref{thm:hardness} suggests that the approximation factor cannot be removed for the case when the target class has treewidth $t \geq 2$,
we are in a nicer landscape of where we only care about $t=1$
case.
Since we already settled to solve the simultaneous version of the problem,
it is enough to find procedures that guarantee to find small sets that are guaranteed to contain a solution variable, in which case we can guess a solution variable $x$
and create a new instance where the number of simultaneous formulas we need to solve double up: by setting $x=0$ in each of the input formulas, and setting $x=1$ in each of the input formulas.
Since we are looking for a solution of size at most $k$,
the number of formulas in a simultaneous instance at any point is bounded by $2^{O(k)}$.

\paragraph{Internal and external cycle killers.}
Coming to our methods for the proof of Theorem~\ref{thm:main}, as we mentioned earlier we will solve the more general simultaneous version of the problem where the instance $(\phi_1, \ldots, \phi_{\alpha},k)$ contains $\alpha \leq 2^{O(k)}$ many formulas.

We say that a variable $x$ kills a cycle $C$ \emph{internally}
if its variable vertex lies on $C$.
It kills $C$ \emph{externally} if it lies outside $C$
and occurs positively in one clause of $C$ and
negatively in another.
In either case, assigning either value to $x$
deletes a vertex of $C$ in the reduct.

With this observation, one is tempted to conclude that our problem
reduces to finding a set of at most $k$ variables that kills every cycle in each input incidence graph, either internally or externally. 
But this is incorrect in general as an assignment can make two clauses identical, causing their clause vertices to be merged.
Together with clause satisfaction, this can destroy a cycle under every assignment to the chosen variables,
even though none of them kills that cycle internally or externally.

\paragraph{The role of a four-cycle $K_{2,2}$.}
To resolve the concern above, we first identify
when clause merging can actually destroy a cycle.
For this to happen, this requires a four-cycle in the original incidence
graph.
Indeed, suppose two distinct clauses become identical
after an assignment and still contain two distinct
variables $u$ and $v$.
Both original clauses must contain $u$ and $v$,
so their clause vertices together with $u,v$ form
a four-cycle.
Thus, if the incidence graph is $K_{2,2}$-free,
only clause vertices of degree at most one can
be merged.
Such merging does not affect acyclicity.

Consequently, when all input incidence graphs are
$K_{2,2}$-free, our cycle-killing viewpoint is exact:
we seek at most $k$ variables that collectively
kill every cycle in every input incidence graph,
internally or externally.
Thus, our first objective is to get rid of cycles of length four in each incidence graph.

\paragraph{Eliminating $K_{2,2}$.}
Eliminating a four-cycle when all clause-lengths is bounded (like when the input formulas are $d$-CNF SAT formulas, for a fixed $d \geq 2$)
is straightforward.
Consider a four-cycle $x,D_1,y,D_2,x$ in one of the
input incidence graphs, where $x,y$ are distinct
variable vertices and $D_1,D_2$ are {\em distinct} clause
vertices representing different sets of literals.
Every simultaneous strong forest backdoor must contain
a variable from $\var(D_1)\cup\var(D_2)$.
Otherwise, assigning the backdoor variables leaves
both clauses unchanged and distinct, so the four-cycle
survives.
If every clause has at most $d$ literals for a fixed
constant $d$, this gives at most $2d$ possible variables
to choose.
One can then branch on choosing a variable
$v\in\var(D_1)\cup\var(D_2)$.

For general CNF formulas, however,
$\var(D_1)\cup\var(D_2)$ can be arbitrarily large,
so branching on all its variables no longer
gives an FPT algorithm.
In order to proceed, 
we first identify the ways in which a four-cycle
$x,D_1,y,D_2,x$ can disappear.
Since $D_1$ and $D_2$ are distinct sets of literals,
$D_1\triangle D_2\neq\emptyset$.
Choose a literal $\ell_0\in D_1\triangle D_2$
and let $z$ be its variable.
We branch on including $x$, $y$, or $z$ in the solution.
If none is chosen, both cycle variables and the
literal distinguishing the clauses remain.
A strong forest backdoor must then contain a variable
occurring with opposite signs in $D_1$ and $D_2$.
Otherwise, an assignment falsifying all chosen-variable
occurrences in these clauses would preserve the four-cycle.
Such an opposite-sign variable satisfies one of
$D_1,D_2$ under either truth value.

There may be arbitrarily many such variables.
So instead of branching on such variables, we instead maintain $k$ sets of variables, alongside our simultaneous instance, $P_1, \ldots, P_k$ that we call {\em candidate sets}.
The idea is to think of the solution strong backdoor as a $k$-sized ordered set whose first vertex comes from $P_1$, second vertex comes from $P_2$ and so on.
Initially, each $P_i$ contains all the variables.

Coming back to the branching strategy, in order to ensure that some variable of the clause $D_1$ or $D_2$ is picked by the solution, instead of branching on these variables, we guess whether this variable is the 1st solution variable, 2nd one and so on: $k$ choices.
In the $i$-th branch, to incorporate this choice we update the $i$-th candidate set
$P_i:= P_i \cap \var(D_1)$ or $P_i: = P_i \cap \var(D_2)$, depending on guessing whether the solution intersects the clause $D_1$ or $D_2$, where $\var(D_i)$ denote the set of variables of clause $D_i$.
The introduction of these candidate sets, useful for controlling such branching leads to defining a more general problem which we call {\sc Simultaneous Colored Strong Forest Backdoor}.

These restrictions record that a chosen variable
occurs in a clause, but do not yet allow us to delete
that clause: whether it is satisfied depends on the
value assigned to the chosen variable.
We therefore refine the branching rule to account
for these values.

For each input formula, we maintain a labelled copy
for every $\sigma\in\{0,1\}^k$.
The coordinate $\sigma_j$ specifies the value assigned
to the variable eventually chosen from $P_j$.
Consider a four-cycle in a copy labelled $\sigma$.
For each position $j$ and clause $E\in\{D_1,D_2\}$,
create a branch in which
\[
    P_j:=
    \{v\in P_j:
      \text{assigning $v=\sigma_j$ satisfies $E$}\}.
\]
Discard the branch if this set is empty.
Otherwise, delete $E$ from that labelled copy,
leaving every other copy unchanged.
This refines the membership restriction described
above and gives only $2k$ branches.

To see why deleting $E$ is safe, consider any eventual
choices $x_j\in P_j$ and any assignment $\tau$ to
their variables.
The matching label is given by
$\sigma_j=\tau(x_j)$ for every $j$.
In this copy, the restriction on $P_j$ guarantees
that $E$ is satisfied and would disappear in the
reduct anyway.
Thus, for every choice and assignment, a matching
copy continues to represent the correct reduct
of each input formula.
Since all copies are obtained only by deleting
clauses, the construction preserves the validity
of every retained choice.

It remains to bound the number of clause deletions
before a variable is chosen.
For this, we prove that every yes-instance has
incidence graphs of {\em degeneracy at most
$\delta:=3^k+k$} (we explain this later).
We reject if any incidence graph exceeds this bound.
Otherwise, each deleted clause contains the entire
candidate set restricted in that branch.
As candidate sets only shrink, this containment
continues to hold at subsequent steps.

While a candidate set has at least $\delta+1$
variables, at most $\delta$ clauses of any original
formula can contain it.
Otherwise, these clauses and $\delta+1$ of those
variables form a $K_{\delta+1,\delta+1}$, contradicting
$\delta$-degeneracy.
Consequently, for each labelled copy and candidate-set
index, only boundedly many deleted clauses can be
counted in this way.
Every deletion adds a new contribution, so the number
of deletion steps along a branch is parameter-bounded.

When some candidate set has at most $\delta$ variables,
we branch on its possible choices, include the chosen
variable in the solution, decrease the budget, and
retain both reduced formulas of every formula.
Thus the procedure either reaches instances whose
incidence graphs are all $K_{2,2}$-free or makes
progress by choosing a solution variable.
This allows us to continue with the ``cycle-killer'' approach.

\paragraph{Win/win based on cyclomatic number.}
After eliminating the four-cycles, we know that the solutions to our problem are precisely the ``cycle-killers'' and vice-versa.
Our strategy going further is a {\em win/win strategy based on the cyclomatic number} of the instances in the simultaneous problem.
The cyclomatic number of a graph $G$,
denoted $\cyc(G)$,
is the minimum number of edges whose deletion results in a forest (note that this is not the feedback vertex set number, but the feedback edge set number).
Thus for a connected graph, its cyclomatic number is the number of edges of the graph minus the number of its vertices plus one.
In general, for any graph $G$, $\cyc(G) = |E(G)| - |V(G)| + \cc(G)$, where $\cc(G)$ denotes the number of connected components of $G$.
A feature of graphs with low cyclomatic number is that they don't have unbounded number of distinct cycles.
In particular, if a graph has cyclomatic number at most $c$, then the number of distinct cycles it has is at most $2^c-1$ (see Lemma~\ref{lem:bounded-number-of-cycles} for the proof).

Let $I=(\Phi=(\phi_1, \ldots, \phi_{\alpha}), V, P_1, \ldots, P_k)$ 
be an instance of {\sc Simultaneous Colored Strong Forest Backdoor} problem, where $V$ is the common variable set of all formulas in $\Phi$ and $P_1, \ldots, P_k$ are the candidate sets which we mentioned earlier while eliminating four-cycles.

    \subparagraph{Small cyclomatic number.}
    Suppose $\cyc(\inc(\phi_i))\leq c  =O(k^3)$ for every $i\in[\alpha]$.
     Incidence graph of each formula in the instance then contains at most $2^{O(k^3)}$ simple cycles.
    Thus, across all formulas, there are  $\alpha 2^{O(k^3)}$ cycles, and $\alpha = 2^{O(k)}$ as we noted earlier.

    We regard these cycles as the elements of a
\textsc{Set Cover} instance.
Each variable $v\in V$ provides a set consisting
of all cycles it kills, internally or externally.
Since the incidence graphs are $K_{2,2}$-free,
covering every cycle is equivalent to choosing
a simultaneous strong forest backdoor.
The candidate-set constraints require us to choose
one variable from each $P_j$.

Without the annotated constraints put by the candidate sets, this is really now just an instance of the {\sc Set Cover} problem which can be solved in time $2^{O(|U|)} \cdot |I|$, where $U$ is the universe of the {\sc Set Cover} instance~\cite{FominKW2004}. 
Since in our case the universe is the set of all distinct cycles,
we can solve such a {\sc Set Cover} instance and thus our problem, in time $2^{\alpha 2^{O(k^3)}} \cdot |I|$.
Thus this gives us an FPT algorithm, except that the running time does not match the one promised in Theorem~\ref{thm:main}.
Before moving forward, note that even with the annotated constraints posed by the candidate sets, one can still mimic the dynamic programming algorithm that solves {\sc Set Cover} in $2^{O(|U|)} \cdot |I|$ time that incorporates these constraints and therefore one can actually solve the full problem in $2^{\alpha 2^{O(k^3)}} \cdot |I|$ time.

To obtain the claimed dependence on $k$ in the running time, we avoid
enumerating all cycles.
Instead, we first partition the variables of each formula
into $2^{O(c)} = 2^{O(k^3)}$ equivalence classes whose members kill exactly the same cycles (see the proof for the details of how to get this partition).
We combine these partitions, across all formulas,
by placing two variables in the same class if and only if they belong to the
same class in the partition for every input formula (that is, we refine these partitions to get one partition for the instance).
Since each formula contributes at most $2^{O(c)}$
classes, the resulting partition has at most
$2^{O(\alpha c)}$ classes.

Thus, variables in the same class kill exactly the
same cycles in every input incidence graph.
For each candidate set $P_j$, we retain one arbitrary
representative from its intersection with each class.
It therefore suffices to try every combination of
one retained representative from each candidate set.
There are at most $\bigl(2^{O(\alpha c)}\bigr)^k
=2^{O(\alpha k c)}$ such combinations.
Since $\alpha=2^{O(k)}$,
this gives the desired double-exponential bound.

\subparagraph{Large cyclomatic number.}
When the cyclomatic number of the incidence graph of some formula is large, we identify
a bounded set of variables around the highest-degree
vertices that every solution must intersect.
Consider the $2$-core of $\inc(\phi)$, which is the graph obtained from $\inc(\phi)$ by repeatedly deleting vertices of degree at most 1. Choose
its $4k$ highest-degree vertices.
Let $S$ contain the chosen variable vertices and
all variables adjacent to at least two chosen
clause vertices.
The first group contains at most $4k$ variables.
Since the incidence graph is $K_{2,2}$-free,
each pair of chosen clause vertices has at most
one common variable neighbor.
Hence $|S|\leq4k+\binom{4k}{2}=O(k^2)$.

We claim that every strong backdoor, in this case a ``cycle-killer'' must intersect $S$ (in which case one can branch on $S$).
Suppose a strong forest backdoor $X$ of size at
most $k$ avoids $S$.
Then $X$ contains no chosen vertex, and each
variable of $X$ is adjacent to at most one chosen
clause vertex.
Thus these high-degree vertices are largely
protected from the variables in $X$.
Moreover, every variable in $X$ has degree at most
that of every chosen vertex.
Counting edges using the requirement that every
assignment to $X$ leaves a forest then bounds
the cyclomatic number by $O(k^3)$.
Above this threshold, avoiding $S$ is therefore
impossible.

\paragraph{Large degeneracy implies a no-instance.}
We show that every non-tautological CNF formula
admitting a strong forest backdoor of size at most
$k$ has incidence degeneracy at most
$\delta:=3^k+k$, which is needed to bound the measure of the branching algorithm while eliminating $K_{2,2}$.

Fix a strong forest backdoor $X$
of size at most $k$.
Each clause has one of at most $3^k$ signed patterns
on $X$: 
each variable occurs positively, negatively,
or not at all.
For each pattern, there is an assignment to $X$ that falsifies
all its literals, preserving every clause with
that pattern.
These clauses remain distinct after their literals
on $X$ are removed.
Their incidences with the remaining variables must
therefore form a forest.
Counting edges across these pattern classes yields
the degeneracy bound.

This is in sharp contrast to weak backdoors:
even a weak forest backdoor of size one places
no bound on the degeneracy of the incidence graph.
Indeed, a variable occurring positively in every
clause satisfies the entire formula when set to true,
while the original incidence graph can have
arbitrarily large degeneracy.
For a strong backdoor, we must also account for
the assignment setting that variable to false.

\paragraph{The role of non-tautologicality.}
The non-tautological assumption is essential to
the degeneracy argument.
It ensures that the literals of a clause on the
backdoor variables can all be falsified by a
suitable assignment.
A tautological clause containing both $x$ and
$\neg x$ cannot be preserved in this way.

Indeed, allowing tautological clauses permits
strong forest backdoors of size one even when
the incidence degeneracy is arbitrarily large.
For example, consider
\[
    \phi_n=
    \bigl\{
        \{x,\neg x,y_1,\ldots,y_n,z_i\}
        :i\in[n]
    \bigr\}.
\]
Assigning either value to $x$ satisfies every
clause, so $\{x\}$ is a strong forest backdoor.
However, the $n$ clause vertices and the variables
$y_1,\ldots,y_n$ induce a $K_{n,n}$,
so the incidence degeneracy is at least $n$.

\subsection{Lower bounds}

\paragraph{The double-exponential dependence.}
We first reduce a $3$-SAT instance on $n$ variables to
\textsc{Set Cover}. Partition the variables into $r$ blocks
of size at most $\lceil n/r\rceil$. The universe has one
element per clause and one per block. Each assignment to
a block gives a set containing that block's element and
the elements of the clauses it satisfies. A cover of size
at most $r$ must choose exactly one assignment per block.
These assignments combine into a satisfying assignment
precisely when their sets cover all clause elements.

We then introduce one variable per set and one incidence
cycle per universe element. The cycle for an element
contains the variables of all sets that cover it. Since
these cycles can share variables, we use selector variables
to expose one element cycle at a time. When the selector
assignment encodes an element, the clauses along its cycle
reduce to binary clauses and the cycle survives. On every
other cycle, the assignment deletes a clause on each path
between consecutive set variables. The resulting pieces
are trees that meet the rest of the incidence graph in at
most one vertex each.

Private cycles force every strong backdoor within the budget
to contain all selectors. The remaining variables must hit
each element cycle under the assignment that exposes it.
A chosen set variable contributes its corresponding set to
the cover, and a chosen variable private to an element cycle
can be replaced by any set covering that element. Conversely,
the selectors together with the variables of a set cover
form a strong forest backdoor, since the chosen set variables
break the exposed cycle under every selector assignment
that exposes one.

There are at most $r2^{\lceil n/r\rceil}$ sets and the
universe has polynomial size in $n$, so we need only
$O(\log n)$ selectors. The resulting formula $\phi$ has
size $2^{O(n/r+\log n)}$ and backdoor budget
$k=r+O(\log n)$. For $r=c\lceil\log n\rceil$ with any fixed
positive integer $c$, a $2^{2^{o(k)}}|\phi|^{O(1)}$-time
backdoor algorithm would decide the original $3$-SAT
instance in time
$2^{(C/c+o(1))n/\log n}$,
where $C$ is independent of $c$. Indeed, for fixed $c$,
we have $k=O(\log n)$, so the inner exponential is
$n^{o(1)}=o(n/\log n)$.

We interleave the runs for $c=1,\ldots,\lceil\log n\rceil$
and return the first answer, since every run decides the
same instance. For each fixed $c$, its run is included for
all sufficiently large $n$. Since $C/c$ can be made
arbitrarily small and the interleaving adds only a polynomial
overhead, this gives a single $2^{o(n/\log n)}$-time
algorithm for $3$-SAT.

\paragraph{Hardness for target treewidth at least two.}
The reduction proceeds in two steps. We first reduce
\textsc{Paired Cut} into \textsc{Simultaneous Treewidth-$t$ Deletion}. We then reduce \textsc{Simultaneous Treewidth-$t$ Deletion} to \strongbackt, for $t \geq 2$.

We start from the \textsc{Paired Cut} problem, defined below, which is \Wone-hard parameterized by $\ell$,
even when the two input graphs have treewidth two. 
In this problem, there are two graphs,
each with two designated terminals, and a list of
selectable pairs of edges, each consisting of one edge from
each graph. The goal is to choose at most $\ell$ pairs
whose deletion disconnects the terminals in both graphs.
We subdivide the two edges of each pair with a common
vertex, so deleting that vertex implements the paired
choice.
We make $\ell+1$ copies of each input graph. Each copy
is a separate graph in the family. The choice vertex for
a given edge pair has the same label in every copy,
whereas every private vertex has a label unique to its
copy. We add missing labels as isolated vertices so that
all graphs have the same vertex set. A deletion set of
size at most $\ell$ therefore leaves at least one copy of
each input graph in which it deletes no private vertex.

At the terminals of each copy, we attach a $K_{t+2}$
with the edge between the terminals missing.
If a path between the terminals survives in the network,
contracting all but one of its edges yields a $K_{t+2}$
minor, so the resulting graph has treewidth greater than $t$.
If the terminals are separated in the remaining network,
each component of that network meets the added gadget
in at most one vertex.
The input networks have treewidth at most two and the added
graph has treewidth $t$, so the resulting graph has
treewidth at most $t$ for $t\geq2$. Thus the \textsc{Paired Cut} instance has a solution of
size at most $\ell$ if and only if one set of at most
$\ell$ vertices reduces the treewidth of every graph
in the constructed family to at most $t$.
Write $G_0,\ldots,G_{\alpha-1}$ for the resulting graphs,
where $\alpha=2(\ell+1)$.
This proves
\Wone-hardness of \textsc{Simultaneous Treewidth-$t$ Deletion}
parameterized by the deletion budget plus the number of graphs.

We then reduce \textsc{Simultaneous Treewidth-$t$ Deletion}
to \strongback\ by encoding the graphs
$G_0,\ldots,G_{\alpha-1}$, where $\alpha=2(\ell+1)$,
in a single non-tautological
$3$-CNF formula using $d=\lceil\log_2\alpha\rceil$ shared selector variables.
Each edge is represented by a chain of clauses joining its endpoint variables. For each $a \in \{0,1, \ldots, \alpha-1\}$, 
assigning the selectors  according to the $d$-bit binary representation of $a$ preserves the chains for $G_a$ as incidence paths. Every other chain breaks into pieces, each containing at most one of its original endpoints. The edge encoding therefore leaves a subdivision of $G_a$ together with trees that meet it in at most one vertex each.

The backdoor budget is $\ell+d$. If $D$ is a common
vertex-deletion set of size at most $\ell$, then $D$
together with the $d$ selectors forms a strong backdoor.
Under the selector assignment for $G_a$, the reduct's
incidence graph consists of a subdivision of $G_a-D$
and trees that meet it in at most one vertex each.
An unused selector assignment leaves a forest.

Conversely, let $X$ be a strong backdoor of size at most
$\ell+d$. The private gadgets force $X$ to contain all
$d$ selectors. From the remaining variables, construct
$D$ by retaining each chosen graph vertex and replacing
each chosen internal chain variable by one endpoint of
its edge. Then $|D|\leq\ell$. For each $a$, every edge of
$G_a-D$ has an incidence path that survives in the reduct
under the selector assignment for $G_a$, since neither
endpoint nor any internal variable of that chain belongs
to $X$. Thus $G_a-D$ is a minor of the reduct's incidence
graph and has treewidth at most $t$.

Since $d=\lceil\log_2(2(\ell+1))\rceil$, the backdoor
budget $\ell+d$ depends only on $\ell$. This establishes
\Wone-hardness of strong backdoor detection for every
fixed $t\geq2$, even on non-tautological $3$-CNF formulas.

\section{Strong Backdoor to Forests is \FPT}
\label{sec:algo}

In this section, we prove \cref{thm:main}.

\strongforestfpt*

To prove \cref{thm:main} we solve a more general problem that asks to find a strong forest backdoor for a given set of formulas simultaneously,
and where the choice of each strong backdoor variable is restricted to come from a specified set of variables.
We define this more general problem next.

\defparproblem
    {Simultaneous Colored Strong Backdoor to Forests}
    {A tuple $\Phi=(\phi_1,\ldots,\phi_\alpha)$ of 
     non-tautological CNF formulas over a common variable set $V$,
     and possibly overlapping sets $P_1,\ldots,P_k\subseteq V$,
     where $\alpha,k\geq1$.}
    {$\alpha + k$}
    {Does there exist $x_j\in P_j$, $j\in[k]$, such that
     $X=\{x_1,\ldots,x_k\}$ is a strong forest backdoor
     for every $\phi_i$, $i\in[\alpha]$?}

The sets $P_1,\ldots,P_k$ are called the
\emph{candidate sets}.
For an instance $I=(\Phi,V,P_1,\ldots,P_k)$ of
\textsc{Simultaneous Colored Strong Backdoor to Forests},
define the size of its instance as
\[
|I|:=|V|+k+\alpha
+\sum_{i=1}^{\alpha}|\phi_i|
+\sum_{j=1}^{k}|P_j|.
\]

\begin{restatable}[Bounded degeneracy]{lemma}{boundeddegeneracy}
\label{lem:bounding-degeneracy-of-yes}
Let $\phi$ be a non-tautological CNF formula and
let $k\geq0$ be an integer.
If $\phi$ has a strong forest backdoor of size at
most $k$, then $\inc(\phi)$ is $(3^k+k)$-degenerate.
\end{restatable}

\begin{proof}
Let $X\subseteq\var(\phi)$ be a strong forest
backdoor with $|X|\leq k$.
It suffices to show that every non-empty induced
subgraph $H$ of $\inc(\phi)$ contains a vertex
of degree at most $3^k+k$.
Put $Y:=(\var(\phi)\cap V(H))\setminus X$ and $ B:=\cla(\phi)\cap V(H)$.
If $B=\emptyset$, then $H$ has no edges.
Hence assume $B\neq\emptyset$.

For each $D\in B$, define its signed pattern
$\sigma_X(D):X\to\{\texttt{0},\texttt{+},\texttt{-}\}$ as
\[
    \sigma_X(D)(x)=
    \begin{cases}
        \texttt{+},&x\in D,\\
        \texttt{-},&\neg x\in D,\\
        \texttt{0},&\text{otherwise}.
    \end{cases}
\]
This is well-defined because $\phi$ is non-tautological.
Group the clauses of $B$ by their signed patterns.
Write $B_1,\ldots,B_t$ for the nonempty groups and
$\pi_1,\ldots,\pi_t$ for their respective patterns.
Then
$1\leq t\leq3^{|X|}\leq3^k$.

For each $i\in[t]$, choose an assignment $\tau_i$
to $X$ that falsifies every literal specified by
$\pi_i$.
No clause of $B_i$ is satisfied by $\tau_i$.
Moreover, distinct clauses in $B_i$ remain distinct
after their literals on $X$ are removed:
otherwise, their identical signed patterns would
imply that the original clauses were identical.
Thus every cycle of $H[Y\cup B_i]$ would survive
in $\inc(\phi[\tau_i])$, contradicting that $X$
is a strong forest backdoor.
Hence $H[Y\cup B_i]$ is a forest and
$ |E(H[Y\cup B_i])|
    \leq |Y|+|B_i|-1$.
Summing over $i\in[t]$ gives
$  |E(H[Y\cup B])|
    \leq t(|Y|-1)+|B|$.

If $|Y|>|B|$ then $ |E(H[Y\cup B])|<(t+1)|Y|$.
Some $y\in Y$ therefore has at most $t$ neighbors
in $B$, which are all its neighbors in $H$.
Thus $\deg_H(y)\leq t\leq3^k$.

Otherwise, $|Y|\leq|B|$, and
$ |E(H[Y\cup B])|
    \leq t(|B|-1)+|B|
    <(t+1)|B|$.
Some clause vertex $D\in B$ therefore has at most
$t$ neighbors in $Y$.
Every other neighbor of $D$ in $H$ belongs to $X$,
so degree of $D$ in $H$ is at most
$t+|X|\leq3^k+k$. This proves the lemma.
\end{proof}

\begin{restatable}{theorem}{coloredboundeddegfpt}
\label{thm:bounded-degeneracy}
Let $d\geq1$ be an integer and let $I=(\Phi,V,P_1,\ldots,P_k)$
be an instance of
\textsc{Simultaneous Colored Strong Backdoor to Forests}.
If $\inc(\phi_i)$ is $d$-degenerate for every $i\in[\alpha]$,
then $I$ can be solved in
$2^{\alpha d \log(d) 2^{O(k)}} \cdot |I|$
time.
\end{restatable}

We first derive \cref{thm:main} from
\cref{thm:bounded-degeneracy}, whose proof occupies
the remainder of this section.

\begin{proof}[Proof of \cref{thm:main}]
Let $(\phi,k)$ be an instance of
\textsc{Strong Backdoor to Forests}.
If $\inc(\phi)$ is a forest, we report a yes-instance.
Otherwise, if $k=0$, we return a no-instance.
These cases can be handled in $O(|\phi|)$ time.
Hence assume $k\geq1$ and that $\inc(\phi)$ contains a cycle.

Set $d:=3^k+k$.
By \cref{lem:bounding-degeneracy-of-yes}, if $(\phi,k)$ is a yes-instance then the degeneracy of $\inc(\phi)$
 is at most $d$.
Using the algorithm of Matula and Beck
\cite{DBLP:journals/jacm/MatulaB83}, we compute the degeneracy
of $\inc(\phi)$ in $O(|\phi|)$ time and return no
if it exceeds $d$.

Otherwise, construct the instance 
$I=(\Phi,V,P_1,\ldots,P_k)$ of {\sc Simultaneous Colored Strong Backdoor to Forests} by setting
$\Phi=(\phi)$, $V=\var(\phi)$, and $P_j=V$ for every $j\in[k]$.
Since $\inc(\phi)$ contains a cycle, every strong forest
backdoor of $\phi$ is nonempty.
Thus any such backdoor of size at most $k$ can be represented
by $k$ choices from $V$, allowing repetitions.
Conversely, every solution to $I$ gives a strong forest
backdoor of $\phi$ of size at most $k$.
The two instances are therefore equivalent.
Also $|I|=
    O((k+1)|\phi|)$.
Applying \cref{thm:bounded-degeneracy}, we solve $I$ in
\[
    2^{d\log d\,2^{O(k)}}\cdot|I|
    =
    2^{2^{O(k)}}\cdot|\phi|
\]
time.
The degeneracy test and construction of $I$ fit within
the same bound, proving the theorem.
\end{proof}

In the remaining section we prove \cref{thm:bounded-degeneracy}.

\subsection{The $K_{2,2}$-free case}\label{sec:k22-free-case}

\paragraph{Cycle killers.}

\begin{definition}[Cycle killer]
Let $\psi$ be a non-tautological CNF formula,
and let $C$ be a simple cycle of $\inc(\psi)$.
A variable is an \emph{internal killer} of $C$
if its variable vertex belongs to $C$.
A variable outside $C$ is an \emph{external killer}
of $C$ if it occurs positively in one clause vertex
of $C$ and negatively in another.
Let $K_\psi(C)$ denote the set of all internal
and external killers of $C$.
\end{definition}

The following lemma characterizes strong forest
backdoors by the cycles they kill.

\begin{restatable}[Cycle-killer characterization]
{lemma}{cyclekiller}
\label{lem:cycle_killer}
Let $\psi$ be a non-tautological CNF formula whose
incidence graph is $K_{2,2}$-free.
Then $X\subseteq\var(\psi)$ is a strong forest
backdoor of $\psi$ if and only if
  $X\cap K_\psi(C)\neq\emptyset$ for every simple cycle $C$ of $\inc(\psi)$.
\end{restatable}

\begin{proof}
First, observe that merging identical clauses preserves the cyclomatic
number, under the stated assumptions.
Indeed, two distinct clauses cannot yield identical
 clauses in a reduct containing two variables:
their original clause vertices would have two
common variable neighbors, forming a $K_{2,2}$.
Thus only isolated clause vertices or clause leaves
with the same variable neighbor can be merged.
These identifications, and the removal of isolated
variable vertices, preserve the cyclomatic number.
Consequently, for every assignment $\tau$ to $X$,
$\cyc(\inc(\psi\semireduct{\tau}))
    =\cyc(\inc(\psi[\tau]))$.

Fix a simple cycle $C$ of $\inc(\psi)$.
If $X$ contains an internal killer of $C$, that
variable vertex is deleted under every assignment.
If $X$ contains an external killer, every assignment
satisfies at least one clause of $C$.
In either case, $C$ survives no assignment to $X$.

Conversely, suppose $X\cap K_\psi(C)=\emptyset$.
No variable vertex of $C$ belongs to $X$.
Moreover, every variable of $X$ occurring among
the clauses of $C$ occurs there with only one
polarity (either only positively or only negatively), because there are no tautological clauses, and there is an  absence
of an external killer in $X$.
Assign these variables to falsify their occurrences
in those clauses, and assign the remaining variables
of $X$ arbitrarily.
Every vertex and edge of $C$ then survives in the
semi-reduct.

Since every semi-reduct incidence graph is a subgraph
of $\inc(\psi)$, it is a forest under every assignment
if and only if $X$ intersects the killer set of every
simple cycle.
A graph is a forest if and only if its cyclomatic number
is zero. By the equality above, for each assignment $\tau$
to $X$, the incidence graph of the semi-reduct is a forest
if and only if the incidence graph of the reduct is a forest.
Thus $X$ is a strong forest backdoor of $\psi$ if and only if
it intersects the killer set of every simple cycle.
\end{proof}

Note that the $K_{2,2}$-free assumption is necessary for Lemma~\ref{lem:cycle_killer}.
For
$ \psi=(a\vee b)\wedge(a\vee b\vee x)$,
both reducts $\psi[x=0]$ and $\psi[x=1]$ consist
of the single clause $a\vee b$.
Thus $\{x\}$ is a strong forest backdoor.
However, the four-cycle through the two clause
vertices and $a,b$ has killer set $\{a,b\}$.
Clause merging can therefore destroy a cycle
without choosing one of its killers.

\paragraph{Restricting backdoors to the $2$-core.}

\begin{restatable}[Restricting backdoors to the $2$-core]
{lemma}{coretransfer}
\label{lem:core-transfer}
Let $\phi$ be a non-tautological CNF formula whose
incidence graph is $K_{2,2}$-free, and put
$ A:=\var(\phi)\cap V(\core_2(\inc(\phi)))$.
For every $X\subseteq\var(\phi)$, $X$ is a strong
forest backdoor of $\phi$ if and only if $X\cap A$
is a strong forest backdoor of $\phi$.
\end{restatable}

\begin{proof}
For every simple cycle $C$ of $\inc(\phi)$, each variable
in $K_\phi(C)$ belongs to the $2$-core of $\inc(\phi)$.
An internal killer lies on the cycle.
An external killer has two distinct clause neighbors
on the cycle; a path between them along the cycle,
together with their edges to the variable, forms
a cycle containing that variable.
Thus every cycle-killer set is contained in $A$.
The equivalence follows from
\cref{lem:cycle_killer}.
\end{proof}

\begin{restatable}[$K_{2,2}$-free case]{lemma}{colorfullinearfpt}
\label{lem:colorful-linear}
Let $I=(\Phi,V,P_1,\ldots,P_k)$ be an instance of
\textsc{Simultaneous Colored Strong Backdoor to Forests},
where $\Phi$ contains $\alpha$ formulas.
If $\inc(\phi_i)$ is $K_{2,2}$-free for every
$i\in[\alpha]$, then $I$ can be solved in
$2^{\alpha2^{O(k)}}\cdot|I|$ time.
\end{restatable}

We prove \cref{lem:colorful-linear} in two cases based on the cyclomatic number of the input formulas.
These two cases are resolved by \cref{lem:low-cyclomatic} and \cref{lem:large-cyclomatic} whose proofs appear in \cref{sec:low-cyclomatic} and \cref{sec:high-cyclomatic}.

\begin{restatable}[Small cyclomatic number]{lemma}{lowcyclomaticfpt}
\label{lem:low-cyclomatic}
Let $c\geq0$ be an integer and let $I=(\Phi,V,P_1,\ldots,P_k)$
be an instance of
\textsc{Simultaneous Colored Strong Backdoor to Forests}, where $\Phi$ contains $\alpha$ formulas.
If $\inc(\phi_i)$ is $K_{2,2}$-free and
$\cyc(\inc(\phi_i))\leq c$ for every $i\in[\alpha]$,
then $I$ can be solved in
$2^{O(\alpha k c)}\cdot|I|$ time.
\end{restatable}

\begin{restatable}[Large cyclomatic number]{lemma}{largecyclomaticbranching}
\label{lem:large-cyclomatic}
For any $k \geq 1$, let $c^*(k)=9k^3 - 3k^2 -k$.
Let $I=(\Phi,V,P_1,\ldots,P_k)$ be an instance of
\textsc{Simultaneous Colored Strong Backdoor to Forests}
such that $\inc(\phi_j)$ is $K_{2,2}$-free for every
$j\in[\alpha]$.
If there exists $i\in[\alpha]$ such that
$\cyc(\inc(\phi_i))\geq c^*(k) +1$, 
then one can compute, in
$O(|\phi_i|)$ time, a set
$S\subseteq\var(\phi_i)$ of size at most $s^*(k) = 2k(4k+1)$
that intersects every strong forest backdoor of $\phi_i$
of size at most $k$.
\end{restatable}

\begin{proof}[Proof of \cref{lem:colorful-linear}]
We give a branching algorithm using
\cref{lem:low-cyclomatic} as the base case and the set $S$ received from
\cref{lem:large-cyclomatic} as a set to branch on.

Let $I'=(\Psi, V, P_1, \ldots, P_{k'})$ be the instance during a recursive call, where $k' \leq k$, and $\Psi$ has $\alpha'$ formulas.
If there exists $j \in [k']$ such that $|P_j| =0$, then report no.
If $k'=0$, report yes if and only if the incidence graph for every formula in $\Psi$
is a forest.
Otherwise, if
$\cyc(\inc(\varphi))\leq c^*(k')$ for every $\varphi\in\Psi$,
we apply \cref{lem:low-cyclomatic} and return its answer.

In the remaining case, choose $\psi\in\Psi$ with
$\cyc(\inc(\psi))>c^*(k')$.
By \cref{lem:large-cyclomatic}, we compute in $O(|\psi|)$
time a set $S\subseteq\var(\psi)$ of size at most
$s^*(k')=2k'(4k'+1)$ intersecting every strong forest backdoor
of $\psi$ of size at most $k'$.
For every $s\in S$ and $j\in[k']$ with $s\in P_j$,
branch on choosing $s$ from $P_j$.
In this branch, remove $P_j$ and replace every formula
$\varphi\in\Psi$ by both formulas $\varphi[s=0]$ and $\varphi[s=1]$.
Keep all other candidate
sets unchanged.
The algorithm reports a yes instance if and only if some branch reports a yes instance.
Note that, each resulting formula is non-tautological and $K_{2,2}$-free, because these properties are preserved
under taking reducts, so both \cref{lem:low-cyclomatic} \cref{lem:large-cyclomatic} remain applicable to the recursive instances.

For correctness, every solution of $I'$ is a strong forest
backdoor of $\psi$ of size at most $k'$, and therefore
intersects $S$ from \cref{lem:large-cyclomatic}.
Thus some branch fixes one of its choices.
Conversely, choices solving a child instance, together with
the fixed choice $s$, solve the parent instance, since
the remaining choices work for both assignments to $s$.
This remains valid when $s$ is also chosen from another
candidate set, since $s$ is absent from both reducts.
Together with the base cases, correctness follows by
induction on $k'$.

Each branching step creates at most
$k'|S|=O(k^3)$ children and decreases the number of
candidate sets by one.
Thus the search tree has depth at most $k$ and
$2^{O(k\log k)}$ nodes.
At depth $r$, there are at most $\alpha2^r$ formulas,
and the instance size is $O(2^r|I|)$.
Processing an internal node, including constructing its
children, takes $k^{O(1)}2^r|I|$ time.
Each leaf handled by \cref{lem:low-cyclomatic}
takes $2^{\alpha2^{O(k)}}|I|$ time, since
$\alpha'\leq\alpha2^k$, $k'\leq k$, and
$c^*(k')=O(k^3)$.
The remaining base cases take linear time in the current
instance size.
Summing over all nodes gives the claimed running time
$2^{\alpha2^{O(k)}}|I|$.\end{proof}

\subsubsection{Small cyclomatic number}\label{sec:low-cyclomatic}

Here we prove \cref{lem:low-cyclomatic}.
By \cref{lem:bounded-number-of-cycles}, the $2$-core
of each incidence graph contains at most $2^c-1$
distinct simple cycles.
We partition the variables of the instance into $2^{O(\alpha c)}$ classes
such that variables in the same class kill
exactly the same set of cycles of the instance. 
We then retain one representative of each such class
inside each candidate set and enumerate the remaining choices.

\lowcyclomaticfpt*

\begin{proof}
If there exists $j \in [k]$ such that $|P_j| =0$, then report a no-instance.
If $c=0$, and for each $\phi \in \Phi$, $\inc(\phi)$ is a forest, report a yes-instance.
Hence assume $c\geq1$.

We first partition $V$ into $2^{O(c)}$ classes such that variables
in the same class kill exactly the same cycles of
$\inc(\phi)$, for a fixed formula $\phi\in\Phi$.

Let $H:=\core_2(\inc(\phi))$.
Recall that for a cycle $C$ in $\inc(\phi)$,
the set $K_\phi(C)$ consists of the variables in $C$
and the variables occurring with both polarities (negatively and positively) among
the clauses of $C$.

\begin{enumerate}
    \item Put all variables of $V$ that are not in $H$ in one class.
    
    Note that every variable that kills a cycle belongs to $H$.
    Indeed, an internal killer lies on the cycle.
    An external killer $x$ has two distinct clause neighbors
    on the cycle; a path between these neighbors along the
    cycle, together with their edges to $x$, forms a cycle
    containing $x$.

    \item Let $B$ be the set of degree at least three vertices of $H$. That is $ B:=\{v\in V(H):\deg_H(v)\geq3\}$.
    Let $t$ be the number of connected components of $H$.
    Since $H$ has minimum degree at least two,
    \[
    |B|
    \leq \sum_{v\in V(H)}(\deg_H(v)-2)
    =2\cyc(H)-2t
    \leq2c.
    \]

    Let $Z$ consist of the variable vertices of $H$ that
    either have degree at least three or have degree two
    and both clause neighbors of degree at least three.
    That is,
    \[
    Z=(B\cap\var(\phi))
      \cup
      \{x\in V(H)\cap\var(\phi):
          \deg_H(x)=2,\ N_H(x)\subseteq B.\}
    \]
    Since $H$ is $K_{2,2}$-free, two degree-two variables in $H$
    cannot have the same pair of neighbors. Therefore,
    $|Z|\leq2c+\binom{2c}{2}$.

    Put each variable of $Z$ in its own class.

    \item Every remaining variable $x$ has degree two in $H$
    and has a clause neighbor of degree two in $H$.
    Such a variable cannot kill a cycle externally:
    otherwise both its clause neighbors would lie on that
    cycle and have degree at least three in $H$.
    Thus $x$ kills precisely the cycles containing it.

    We partition these variables according to the cycles
    containing them, as follows.
    
    Choose a spanning forest $T$ of $H$, and write
    $E(H)\setminus E(T)=\{e_1,\ldots,e_h\}$ where $ h=\cyc(H)\leq c$.
    For each $\ell\in[h]$, let $D_\ell$ be the unique
    cycle in $T+e_\ell$.
    Every simple cycle $C$ of $H$ satisfies
    \begin{equation}\label{eq:sym-diff}
        E(C)= \mathop{\triangle}_{\ell:\,e_\ell\in E(C)}
        E(D_\ell),
    \end{equation}  
        where $\triangle$ denotes symmetric difference.
    Indeed, the symmetric difference of the two sides
    is an even-degree subgraph of the forest $T$,
    and is therefore empty.

     For each remaining variable $x$, choose an incident
    edge $f_x$ and define
    $\lambda(x):=
        \bigl(\mathbf{1}_{f_x\in E(D_\ell)}\bigr)_{\ell\in[h]}$.
         Since $x$ has degree two in $H$, a cycle contains $x$
    if and only if it contains $f_x$.
    Put variables $x$ and $y$ in the same class if
    $\lambda(x)=\lambda(y)$.
Then $f_x$ and $f_y$ belong to exactly the same cycles
$D_\ell$ and, by \cref{eq:sym-diff}, 
to exactly
the same simple cycles of $H$.
Since $x$ and $y$ have degree two, they therefore
belong to, and kill, exactly the same cycles.
There are at most $2^h\leq2^c$ such classes.

\end{enumerate}

Thus the total number of classes in which we partition the variable set $V$ is at most
\[
    M:=1+2c+\binom{2c}{2}+2^c=2^{O(c)}.
\]

The $2$-core can be computed in linear
time~\cite{DBLP:journals/adac/BatageljZ11}.
A spanning forest can then be computed by
depth-first search in linear
time~\cite[Section~22.3]{DBLP:books/mg/CormenLRS01}.
For each edge outside the spanning forest, we use
depth-first search in the forest to find the unique
path between its endpoints and add the edge to form
its fundamental cycle. Each search takes $O(|\phi|)$
time, so constructing all $h\leq c$ fundamental cycles
takes $O(c|\phi|)$ time.
The binary vectors $\lambda(x)$ have at most $c$ coordinates.
Using the integer sorting procedure in \cref{sec:preliminaries},
we group the variables by their vectors $\lambda(x)$ in
$O(c|V|)$ time. Two variables are in the same group exactly
when their vectors agree in every coordinate.
Thus, for the fixed formula $\phi$, we can partition $V$
in $O(c(|\phi|+|V|))$ time so that any two variables in the
same class kill exactly the same cycles of $\inc(\phi)$.

For each $\phi_i\in\Phi$, compute the partition of $V$
described above.
Let $\sigma_i(x)$ denote the class of $x$ for $\phi_i$,
and define the joint signature
\[
    \sigma(x):=(\sigma_1(x),\ldots,\sigma_\alpha(x)).
\]
There are at most $M^\alpha=2^{O(\alpha c)}$
joint signatures.
For each $j\in[k]$, retain a set $R_j\subseteq P_j$
containing one representative of every joint signature
occurring in $P_j$.
Enumerate all tuples
\[
    (x_1,\ldots,x_k)\in R_1\times\cdots\times R_k
\]
and accept if their chosen variables form a strong
forest backdoor for every formula in $\Phi$.

For correctness, since each $\inc(\phi_i)$ is
$K_{2,2}$-free, \cref{lem:cycle_killer} implies
that $X$ is a strong forest backdoor of $\phi_i$
if and only if
$X\cap K_{\phi_i}(C)\ne\emptyset$
for every simple cycle $C$ of $\inc(\phi_i)$.
Replacing any chosen variable by another variable
with the same joint signature preserves all these
intersections in every formula.
Every feasible tuple can therefore be replaced by a
tuple from $R_1\times\cdots\times R_k$.
Conversely, every accepted tuple chooses its $j$th
variable from $P_j$ and is a solution to $I$.
Conversely, every accepted tuple chooses its $j$th
variable from $P_j$ and is a solution to $I$.

There are at most $(M^\alpha)^k=2^{O(\alpha k c)}$ tuples to test.
For each tuple, let $X$ be its set of chosen variables.
Consider all assignments $\tau$ to $X$ and test whether
$\inc(\phi_i[\tau])$ is a forest for every $i\in[\alpha]$.
These tests take $O(2^k|I|)$ time per tuple.
For these tests, clause vertices need not be merged:
$K_{2,2}$-freeness ensures that distinct clauses can
become identical only when they contain at most one
variable in the reduct, and merging such clause
vertices preserves acyclicity.

Assign the classes of each formula's partition
integer labels in $[|V|]$.
The joint signatures have $\alpha$ coordinates,
so the integer sorting procedure in
\cref{sec:preliminaries} groups them in
$O(\alpha|V|)$ time.
Assign the resulting classes integer labels
in $[|V|]$; write $\ell(x) \in [|V|]$ for the label of the class $x$ belongs to.

For a fixed $j \in [k]$, to obtain the representative sets $R_j$,
create a record $(j, \ell(x), x)$ for each $x \in P_j$. Sort these
records by their first two coordinate using the same integer sorting procedure
in \cref{sec:preliminaries}; and then retain one variable per distinct
pair $(j, \ell(x))$ to construct $R_j$. The total time over all $j \in [k]$
this takes is $O(\sum_{j} |P_j| + k + |V|) = O(|I|)$.

Together with constructing the individual partitions,
constructing the representative sets takes
$O(\alpha c|I|)$ time.
Combining this construction time with the
$2^{O(\alpha k c)}$ tuples and the
$O(2^k|I|)$ time per tuple gives total running time
$2^{O(\alpha k c)}|I|$.
\end{proof}

\subsubsection{Large cyclomatic number}\label{sec:high-cyclomatic}

Here we prove \cref{lem:large-cyclomatic}.
Our goal is to compute a small set $S$ that intersects
every strong forest backdoor of $\phi_i$ of size at most
$k$, regardless of the additional constraints of the
simultaneous colored problem.
The proof has three steps.

First, let $H:=\core_2(\inc(\phi_i))$ and choose the $4k$
highest-degree vertices of $H$, with degrees measured
in $H$. These may be variable vertices or clause vertices.
Put into $S$ all chosen variable vertices and every
variable adjacent to at least two chosen clause vertices.
By $K_{2,2}$-freeness, each pair of chosen clause vertices
has at most one common variable neighbor, so $|S|=O(k^2)$.

Second, suppose that a strong forest backdoor $X$ of
size at most $k$ avoids $S$.
We show that every vertex outside the chosen $4k$
vertices has degree at most $3k$ in $H$.
This is the main technical step.
Deleting $X$ and all its neighboring clause vertices
leaves a forest containing at least $3k$ of the chosen
vertices: no chosen variable is deleted, and each
variable of $X$ has at most one chosen clause neighbor.
If the remaining chosen vertices had very large degrees
in $H$, this forest would have too many edges leading
to the deleted clauses.
We obtain a matching upper bound on these edges by
grouping the deleted clauses so that all clauses in
each group can survive together under an assignment
to $X$. Comparing the bounds gives the required
degree bound.

Finally, we use this degree bound to bound $\cyc(H)$.
Since each variable of $X$ has at most one chosen clause
neighbor, non-tautologicality lets us assign $X$ so that
every chosen clause survives.
Thus the assignment deletes only variable and clause
vertices outside the chosen set, whose degrees are
bounded by $3k$.
We bound the decrease in cyclomatic number caused by
these deletions. Since the resulting graph is a forest,
this gives $\cyc(H)\leq c^*(k)$, contrary to the hypothesis.
Consequently, every strong forest backdoor of size at
most $k$ intersects $S$.

\largecyclomaticbranching*

\begin{proof}
    Put $\psi:=\phi_i$ and $H:=\core_2(\inc(\psi))$.
Let $ A:=\var(\psi)\cap V(H)$ and $B:=\cla(\psi)\cap V(H)$.
Then $ \cyc(H)=\cyc(\inc(\psi))>c^*(k)$.
In particular, $H$ is nonempty and has minimum degree
at least two. Set $g:=4k$ and recall that
$c^*(k)=9k^3-3k^2-k$.

\paragraph{Constructing the branching set.}
We first observe that $|V(H)| >g$. Indeed, as otherwise, since $H$ is a bipartite graph, $\cyc(H)\leq |E(H)|  \leq \frac{|V(H)|^2}{4} \leq4k^2\leq c^*(k)$, which is a contradiction.

Let $T \subseteq V(H)$ consist of the $g$ highest-degree vertices of
$H$, breaking ties arbitrarily, and define
$T_A:=T\cap A$, $T_B:=T\cap B$, $T_A':=\{x\in A\setminus T_A: |N_H(x)\cap T_B|\geq2\}$ and $S:=T_A\cup T_A'$.

By $K_{2,2}$-freeness, each pair of clause vertices
has at most one common variable neighbor.
Associating each variable of $T_A'$ with any pair
of its neighbors in $T_B$ therefore gives
$ |S|
    \leq |T_A|+\binom{|T_B|}{2}
    \leq g+\binom g2
    =2k(4k+1)$.

The degrees in $H$ are at most $|V(H)|-1$.
Using counting sort as described in
\cref{sec:preliminaries}, we construct $T$ by
selecting the $4k$ highest-degree vertices in
$O(|V(H)|+|E(H)|)$ time.
Constructing $S$ then requires only a scan of
the adjacency lists.
Thus $S$ is computable in $O(|\psi|)$ time.

Suppose, for a contradiction, that $\psi$ has a strong
forest backdoor $X$ of size at most $k$ with
$X\cap S=\emptyset$.
By \cref{lem:core-transfer}, we may replace $X$ by
$X\cap A$ and hence assume $X\subseteq A$.

\paragraph{Bounding the degrees outside $T$.}
Put $ D:=\max\{\deg_H(v):v\in V(H)\setminus T\}$.
Since $X\cap S=\emptyset$, every variable of $X$
lies outside $T$ and has at most one neighbor in $T_B$.
Thus $\deg_H(x)\leq D$ for every $x\in X$.

For every assignment $\tau$ to $X$, the graph
$\inc(\psi\semireduct{\tau})$ is a forest.
Indeed, $X$ is a strong forest backdoor, and
$K_{2,2}$-freeness ensures that only clause vertices
containing at most one variable can be merged.

Define
$R:=H-N_H[X]$ and $N_H[X]:=X\cup N_H(X)$.
Every vertex and edge of $R$ survives every assignment
to $X$, so $R$ is a forest.
No vertex of $T_A$ is deleted, and at most $k$ vertices
of $T_B$ are deleted. Hence $R$ contains at least
$g-k=3k$ vertices of $T$.

Let $W$ be the union of the components of $R$ that
contain a vertex of $T$. Write
$p:=|T\cap V(W)|\geq3k$,
and let $t$ be the number of components of $W$.
Then $1\leq t\leq p$ and $|E(W)|=|V(W)|-t$.

For disjoint vertex sets $U,U'$, write $E_H(U,U')$
for the set of edges of $H$ between them.
Every edge leaving $W$ ends in $N_H(X)$.
The $p$ vertices of $T\cap V(W)$ have degree at least
$D$ in $H$, and every other vertex of $W$ has degree
at least two. Therefore,
\begin{align}
    |E_H(V(W),N_H(X))|
    &=\sum_{v\in V(W)}\deg_H(v)-2|E(W)|
      \notag\\
    &\geq 2|V(W)|+p(D-2)-2(|V(W)|-t)
      \notag\\
    &=2t+p(D-2).
    \label{eq:lower-degree}
\end{align}
We next obtain an upper bound on the same quantity.
For each clause $C\in N_H(X)$, choose a literal
$\ell_C\in C$ whose variable belongs to $X$.
Group these clauses according to the chosen literal,
obtaining nonempty groups $Q_1,\ldots,Q_h$, where
$h\leq2|X|\leq2k$.

Fix a group $Q_j$, whose chosen literal is $\ell$
on a variable $x\in X$.
By $K_{2,2}$-freeness, any two clauses of $Q_j$
share no variable other than $x$.
We can therefore assign $X$ so that every clause of
$Q_j$ survives: falsify $\ell$ and every occurrence
of a variable of $X\setminus\{x\}$ in these clauses.
This is consistent because each such variable occurs
in at most one clause of $Q_j$, and non-tautologicality
ensures that it occurs there with only one polarity (either only positively or only negatively).

Under this assignment, all vertices and edges of $W$
also survive. Consequently, the graph with vertex set
$V(W)\cup Q_j$ and edge set
$E(W)\cup E_H(V(W),Q_j)$
is a forest. Counting its edges and vertices gives
$|V(W)|-t+|E_H(V(W),Q_j)|
    \leq |V(W)|+|Q_j|$,
and hence
$|E_H(V(W),Q_j)|\leq t+|Q_j|$.
Since
$|N_H(X)|
    \leq\sum_{x\in X}\deg_H(x)\leq kD$,
summing over the groups yields
\begin{equation}\label{eq:upper-degree}
    |E_H(V(W),N_H(X))|
    \leq ht+|N_H(X)|
    \leq2kt+kD.
\end{equation}
Combining \eqref{eq:lower-degree} and
\eqref{eq:upper-degree}, we obtain
$ (p-k)D
    \leq2p+(2k-2)t
    \leq2kp$,
where the last inequality uses $t\leq p$.
Since $p\geq3k$,
$ D\leq\frac{2kp}{p-k}\leq3k$.

\paragraph{Bounding the cyclomatic number.}

Choose an assignment $\tau: X \to \{0,1\}$ 
that falsifies
every occurrence of a variable of $X$ in a clause
of $T_B$.
Such an assignment exists because each variable of
$X$ has at most one neighbor in $T_B$, and
non-tautologicality ensures that it occurs there
with only one polarity (either only positively or only negatively).
Thus no clause of $T_B$ is satisfied by $\tau$.

Deleting a vertex of degree $a$ decreases the
cyclomatic number by at most $\max\{a-1,0\}$.
First delete the variable vertices of $X$ from $H$.
Their degrees are at most $D$, so these deletions
decrease the cyclomatic number by at most $k(D-1)$.

Let $Z_\tau\subseteq B$ consist of the clauses
satisfied by $\tau$.
Each such clause is adjacent to $X$ and lies outside
$T_B$, so
$Z_\tau\subseteq N_H(X)\setminus T_B$, $|Z_\tau|\leq kD$.
Every vertex of $Z_\tau$ lies outside $T$ and hence
has degree at most $D$ in $H$.
It also loses at least one neighbor when $X$ is deleted.
Its remaining degree is therefore at most $D-1$,
so deleting it decreases the cyclomatic number by
at most $D-2$.
Here $D\geq2$, since $H$ has minimum degree at least two.

The graph remaining after deleting $X$ and $Z_\tau$
is a subgraph of $\inc(\psi\semireduct{\tau})$ and
is therefore a forest.
Consequently,
\[
\begin{aligned}
    \cyc(H)
    &\leq k(D-1)+kD(D-2)\\
    &=k(D^2-D-1)\\
    &\leq k(9k^2-3k-1)
     =c^*(k),
\end{aligned}
\]
where the last inequality uses $2\leq D\leq3k$.
This contradicts $\cyc(H)>c^*(k)$.

Thus every strong forest backdoor of $\psi$ of size
at most $k$ intersects $S$.
The set $S$ has size at most $2k(4k+1)$ and is
computable in $O(|\psi|)=O(|\phi_i|)$ time.

\end{proof}

\subsection{Dealing with $K_{2,2}$}\label{subsec:dealing-with-k22}

In this subsection, we reduce the problem to instances
whose formulas have $K_{2,2}$-free incidence graphs.
The main difficulty is that a four-cycle can disappear
not only through deleting a variable or satisfying a
clause, but also through merging two clauses that become
identical after an assignment.

Consider a four-cycle $x,C,y,D,x$ in a current formula.
Since $C$ and $D$ are distinct, choose a literal in
$C\triangle D$ and let $z$ be its variable.
We branch on choosing $x$, $y$, or $z$ from a candidate
set. If none is chosen, both cycle variables and a
literal distinguishing the clauses remain.
Consequently, a strong forest backdoor must contain
a variable occurring with opposite signs in $C$ and $D$.
Under any assignment, this variable satisfies one of
the two clauses.

To preserve the requirement that every assignment to
the backdoor variables leaves a forest incidence graph
for every formula, we maintain copies of the formulas
labelled by Boolean values for the $k$ candidate-set indices.
For a labelled copy, we branch on which candidate set
provides a variable satisfying $C$ or $D$, restrict that
candidate set accordingly, and delete the satisfied
clause from that copy.
The labels allow us to maintain a copy giving the correct
reduct of each original formula under every assignment
to the chosen variables.

Finally, we use $d$-degeneracy to bound the number
of clause-deletion steps along any branch.
Each deletion restricts a candidate set to variables
of the deleted clause.
While every candidate set has at least $d+1$
variables, only boundedly many deleted clauses
can contain an entire candidate set.
Since candidate sets only shrink, these contributions
accumulate throughout the branch.
The procedure stops when all incidence graphs are
$K_{2,2}$-free or outputs instances with a singleton
candidate set.

\paragraph{Finding a four-cycle.}
While proving \cref{lem:four-cycle-preprocessing}, we use the following lemma to find a $K_{2,2}$ in linear time in a graph with bounded degeneracy.
\begin{restatable}[Finding $K_{2,2}$]{lemma}{findktwotwo}
\label{lem:find-k22}
Let $d\geq1$ be an integer and let $G$ be a
$d$-degenerate graph.
In $O(d^4(|V(G)|+|E(G)|))$ time, one can find
a copy of $K_{2,2}$ in $G$ or correctly report
that $G$ is $K_{2,2}$-free.
\end{restatable}

\begin{proof}
Let $n:=|V(G)|$. Using the algorithm of Matula and
Beck~\cite{DBLP:journals/jacm/MatulaB83}, compute in
$O(n+|E(G)|)$ time an ordering $v_1,\ldots,v_n$
in which each vertex has at most $d$ neighbors
that occur after it. For each $i\in[n]$, store these
neighbors of $v_i$ in the list
$ L_i:=\{v_j\in N_G(v_i):j>i\}$.
For $i<j$, the vertices $v_i$ and $v_j$ are adjacent
if and only if $v_j\in L_i$. Since $|L_i|\leq d$,
scanning $L_i$ tests adjacency in $O(d)$ time.

For every $i\in[n]$ and every pair
$v_a,v_b\in L_i$ with $a<b$, create the triple
$(i,a,b)$.
There are $O(nd^2)$ triples.
We group the triples by $(a,b)$ to detect two
distinct vertices sharing the later neighbors
$v_a,v_b$.
Use the integer sorting procedure described in
\cref{sec:preliminaries}, with keys $a,b\in[n]$.
Since there are $O(nd^2)$ triples, this takes
$O(n+nd^2)$ time.

If a group contains distinct triples $(i,a,b)$
and $(i',a,b)$, return the copy of $K_{2,2}$
with bipartition
$ \{v_i,v_{i'}\},\{v_a,v_b\}$.
These four vertices are distinct because
$i,i'<a<b$.

Otherwise, for each triple $(i,a,b)$, examine
every $w\in L_a\setminus\{v_b\}$.
If $w$ is adjacent to $v_b$, return the copy
of $K_{2,2}$ with bipartition
$\{v_i,w\},\{v_a,v_b\}$.
Here $w$ is adjacent to $v_a$ by its membership
in $L_a$.
The four vertices are distinct because $w$ appears
later than $v_a$, while $i<a<b$, and $w\neq v_b$.
If no copy is found, report that $G$ is
$K_{2,2}$-free.

To prove correctness, suppose $G$ contains a cycle
$v_i,v_a,v_r,v_b,v_i$ of length four, where $v_i$
is its earliest vertex.
Interchange $a$ and $b$ if necessary so that $a<b$.
The triple $(i,a,b)$ is created.
If $r<a$, then $(r,a,b)$ is also created, and
the grouping step finds a copy of $K_{2,2}$.
Otherwise, $r>a$, so $v_r\in L_a\setminus\{v_b\}$
and is adjacent to $v_b$.
The second step therefore finds a copy.

Each triple requires examining at most $d$
vertices, with an $O(d)$ adjacency test for each.
The second step takes $O(nd^4)$ time.
Together with constructing the ordering and
grouping the triples, the total running time is
$O(d^4(n+|E(G)|))$.
\end{proof}

\begin{restatable}[Eliminating $K_{2,2}$]{lemma}{fourcyclepreprocessing}
\label{lem:four-cycle-preprocessing}
Let $d \geq2$ be an integer and let $I=(\Phi,V,P_1,\ldots,P_k)$
be an instance of
\textsc{Simultaneous Colored Strong Backdoor to Forests}.
Suppose that $\inc(\phi_i)$ is $d$-degenerate for every
$i\in[\alpha]$.
In $2^{\alpha d\log d\,2^{O(k)}}\cdot|I|$ time, 
one can either

\begin{enumerate}
    \item correctly decide $I$, or
    \item construct a family of at most
$2^{\alpha d\log d\,2^{O(k)}}$ instances
$I_1,\ldots,I_{\ell}$ of the same problem such that:
\begin{enumerate}[label=(\roman*)]
    \item $I$ is a yes-instance if and only if at least one
    $I_s$, $s\in[\ell]$, is a yes-instance.

    \item For every $s\in[\ell]$, $I_s$ has $k$
    candidate sets, at most $\alpha2^k$ formulas, and
    $|I_s|\leq2^{O(k)}\cdot|I|$.

    \item The incidence graph of every formula in each $I_s$
    is $d$-degenerate. Moreover, for each $s \in [\ell]$, either
    \begin{itemize}
        \item the incidence graph of every formula in  $I_s$ is $K_{2,2}$-free, or
        \item $I_s$ has a singleton candidate set
    $\{v_s\}$; in this case, the algorithm also returns
    $v_s$, which belongs to every solution of $I_s$.
    \end{itemize}
    \item Each output instance has the form
$I_s=(\Phi_s,V,P^s_1,\ldots,P^s_k)$, where
$P^s_j\subseteq P_j$ for every $j\in[k]$.
For every choice $x_j\in P^s_j$, the set
$X=\{x_1,\ldots,x_k\}$ is a strong forest backdoor
of every formula in $\Phi_s$ if and only if it
is a strong forest backdoor of every formula
in $\Phi$.
\end{enumerate}
All output instances use the original variable set $V$.
\end{enumerate}

\end{restatable}

\begin{proof}
We give a recursive procedure.
For the proof, write the input instance as
$I_0=(\Phi,V,P^0_1,\ldots,P^0_k)$.
The parameters $\alpha,k,d$ and the variable set
$V$ remain fixed throughout the procedure.

\subparagraph{Instance.}
Every recursive instance has the form
$\widehat I=(\Psi,V,P_1,\ldots,P_k)$, where $P_j\subseteq P^0_j$
for every $j\in[k]$.
The tuple $\Psi$ contains $\alpha2^k$ formulas $\psi_{i,\sigma}$,
indexed by $i\in[\alpha]$ and $\sigma\in\{0,1\}^k$.
Initially, $P_j=P^0_j$ for each $j \in [k]$ and $\psi_{i,\sigma}=\phi_i$
for each $i \in [\alpha]$ and $\sigma \in \{0,1\}^k$.
The number $k$ (of candidate sets) and the variable set $V$ remain unchanged
in every recursive call.

\subparagraph{Invariants.}
In every recursive instance, we maintain the following invariants:
\begin{enumerate}
\item For every $i\in[\alpha]$ and $\sigma\in\{0,1\}^k$,
$\psi_{i,\sigma}$ is a subformula of $\phi_i$,
that is, $\psi_{i,\sigma}\subseteq\phi_i$
as sets of clauses.

\item For every choice $x_j\in P_j$ and every assignment
$\tau:\{x_1,\ldots,x_k\}\to\{0,1\}$, define
$\sigma^\tau_j=\tau(x_j)$ for $j\in[k]$.
Then, for every $i\in[\alpha]$,
\[
 \psi_{i,\sigma^\tau}[\tau]=\phi_i[\tau].
\]
\end{enumerate}

\subparagraph{Base cases.}

\begin{enumerate}
    \item If $P_j=\emptyset$ for some $j\in[k]$,
    reject the branch.

    \item If $1\leq|P_j|\leq d$ for some $j\in[k]$,
    choose one such index and output, for every
    $v\in P_j$, the instance
    \[
        (\Psi,V,P_1,\ldots,P_{j-1},
                 \{v\},P_{j+1},\ldots,P_k).
    \]
    Return $v$ with the corresponding output
    instance and terminate the branch.

    \item If every incidence graph of a formula
    in $\Psi$ is $K_{2,2}$-free, output
    $\widehat I$ and terminate the branch.
\end{enumerate}

Therefore we can now assume that
$|P_j|\geq d+1$ for every $j\in[k]$,
and some formula in $\Psi$ has an incidence graph
containing a $K_{2,2}$.

\subparagraph{Branching.}
If none of the base cases apply,
then there exists a formula, say $\psi_{i,\sigma} \in \Psi$, such that $\inc(\psi_{i,\sigma})$ contains a $K_{2,2}$, that is a cycle on $4$ vertices.
Using \cref{lem:find-k22}, find a four-cycle
$x,C,y,D,x$ in some $\inc(\psi_{i,\sigma})$.
Searching through all formula entries takes
$O\left(
        d^4\sum_{i\in[\alpha]}
           \sum_{\sigma\in\{0,1\}^k}
           (1+|\psi_{i,\sigma}|)
    \right)$ time.
Here $x,y \in V$ and $C,D$ are the (distinct) clauses in $\psi_{i, \sigma}$.
Since $C,D$ are distinct sets of literals, choose
$\ell_0\in C\triangle D$ and let $z$ be its underlying variable.

Create the following branches:
\begin{enumerate}
\item \emph{Choose $x$.}
For every $j\in[k]$ with $x\in P_j$, create the child
\[
 (\Psi,V,P_1,\ldots,P_{j-1},\{x\},P_{j+1},\ldots,P_k).
\]

\item \emph{Choose $y$.}
For every $j\in[k]$ with $y\in P_j$, create the child
\[
 (\Psi,V,P_1,\ldots,P_{j-1},\{y\},P_{j+1},\ldots,P_k).
\]

\item \emph{Choose $z$.}
For every $j\in[k]$ with $z\in P_j$, create the child
\[
 (\Psi,V,P_1,\ldots,P_{j-1},\{z\},P_{j+1},\ldots,P_k).
\]

\item \emph{Satisfy $C$ or $D$ under the label $\sigma$.}
The set $\var(C) \cup \var(D)$ 
may be large, so we do not branch
on its individual variables.
Instead, we guess which candidate set $P_j$ supplies a variable
and which clause $E\in\{C,D\}$ is satisfied by setting
that variable to the value given by $\sigma_j$.

For every $j\in[k]$ and $E\in\{C,D\}$, put
\[
 R_{j,E}
 :=
 \{v\in P_j:\text{assigning $v$ the value given by $\sigma_j$ satisfies $E$}\}.
\]
If $R_{j,E}\neq\emptyset$, create the child
\[
 (\Psi^E,V,P_1,\ldots,P_{j-1},R_{j,E},P_{j+1},\ldots,P_k),
\]
where $\Psi^E$ is obtained by deleting $E$ from
$\psi_{i,\sigma}$, the formula containing the chosen four-cycle.
All other labelled formulas remain unchanged, even if they also
contain $E$.
\end{enumerate}

There are at most $3k+2k=5k$ children.
If no child is created, reject the branch.
Otherwise, recursively process the children and collect their
output instances.

\subparagraph{Correctness.}
We show that, whenever a recursive call creates child instances,
its input has a solution if and only if at least one of those
children has a solution.
The base cases are immediate: an empty candidate set admits no
solution, and the small-sized candidate set case exhaustively considers
all choices from a fixed $P_j$.

{\textit Invariants.} We first prove that both invariants are maintained.
Both hold initially.
Restricting candidate sets preserves them for all choices still allowed.
Deleting clauses preserves the first property.
For the second property, consider a deletion of $E$ from
$\psi_{i,\sigma}$ in the fourth branching rule.
For every retained choice, $x_j\in R_{j,E}$.
If $\sigma^\tau=\sigma$, then $\tau(x_j)=\sigma_j$,
so $E$ is satisfied by $\tau$.
Hence $E$ would be deleted in the reduct anyway, and deleting
it early does not change that reduct.
If $\sigma^\tau\neq\sigma$, the matching labelled formula is unchanged.
All other original indices are unaffected.
Thus both invariants are maintained.

{\textit Backdoor property.}
The invariants imply that, for any current choice $x_j\in P_j$,
the set $X=\{x_1,\ldots,x_k\}$ is a common strong forest backdoor
of $\Phi$ if and only if it is a common strong forest backdoor of $\Psi$.
Indeed, if $X$ is a common strong forest backdoor of $\Phi$,
every reduct of a current formula is a subformula of the
corresponding original reduct, by the first property,
and therefore its incidence graph is a forest.
Conversely, suppose that $X$ is a common strong forest backdoor
of $\Psi$, and let $\tau:X\to\{0,1\}$ be any assignment.
Consider its matching label $\sigma^\tau$.
Since $X$ is a strong forest backdoor of every formula in $\Psi$,
the graph $\inc(\psi_{i,\sigma^\tau}[\tau])$ is a forest
for every $i\in[\alpha]$.
By the second invariant,
$\psi_{i,\sigma^\tau}[\tau]=\phi_i[\tau]$,
so $\inc(\phi_i[\tau])$ is also a forest for every $i\in[\alpha]$.
Since $\tau$ was arbitrary, $X$ is a common strong forest backdoor
of $\Phi$.

We have shown that changing the formulas preserves
whether any fixed choice $x_j\in P_j$, $j\in[k]$,
gives a set $X=\{x_1,\ldots,x_k\}$ that is a strong forest backdoor of every formula.
We now show that, when the candidate sets are restricted, every solution to the current instance
remains a solution to at least one child or output instance.

{\textit Exhaustive.}
The small-candidate-set case considers every possible choice from
the selected $P_j$, while the $K_{2,2}$-free base case outputs the
current instance unchanged.

For four-cycle branching, consider a solution $(x_1,\ldots,x_k)$
with $x_j\in P_j$, and put $X=\{x_1,\ldots,x_k\}$.
If $x_j\in\{x,y,z\}$ for some $j$, the corresponding branch among
the first three kinds fixes $P_j=\{x_j\}$ and retains the entire tuple.

Otherwise, $X\cap\{x,y,z\}=\emptyset$.
We claim that some variable of $X$ occurs with opposite signs
in $C$ and $D$.
If not, non-tautologicality of $\psi_{i,\sigma}$ allows an assignment
$\rho:X\to\{0,1\}$ that falsifies every occurrence of a variable
from $X$ in $C$ or $D$.
Neither clause is satisfied, and both residual clauses still contain
$x,y$.
Since $z\notin X$, the literal $\ell_0$ remains
in exactly one of the two reduct clauses.
Thus these clauses remain distinct after
duplicate clauses are merged.
They therefore form a four-cycle with $x,y$ in
$\inc(\psi_{i,\sigma}[\rho])$, contradicting that $X$ is a strong
forest backdoor of $\psi_{i,\sigma}$.

Choose $j$ such that $x_j$ occurs with opposite signs in $C,D$.
Setting $x_j=\sigma_j$ satisfies exactly one of these clauses,
say $E$.
Thus $x_j\in R_{j,E}$, and the corresponding fourth branch
retains the entire tuple.

Conversely, every solution of a child is an allowed choice
in its parent, because candidate sets only shrink.
By the invariants, its variable set is a common strong forest
backdoor of $\Phi$ and therefore also of the parent's formulas.
Hence a branching instance has a solution if and only if at
least one of its children has a solution.
In particular, if no child is created, the current instance has no
solution, which justifies rejecting the branch.

\emph{Overall.}
Together with termination, proved below,
branching equivalence shows that $I_0$ is a
yes-instance if and only if at least one output
instance is a yes-instance.
If no instance is output, report no.

Every output instance has the form
$I_s=(\Phi_s,V,P^s_1,\ldots,P^s_k)$,
with $P^s_j\subseteq P^0_j$ for every $j\in[k]$.
By the backdoor property proved above, every
choice $x_j\in P^s_j$ gives a strong forest
backdoor of every formula in $\Phi_s$ if and
only if it gives one for every formula in $\Phi$.
This establishes item~(iv).

By the base cases, every output instance either
has only $K_{2,2}$-free incidence graphs or has
a specified singleton candidate set $\{v_s\}$.
In the latter case, every solution contains $v_s$.

\subparagraph{Measure and depth.}
Consider the recursion tree, whose nodes correspond to recursive
instances and whose edges correspond to recursive calls.
A node is a leaf if the procedure stops at that instance, either
because a base case applies or because no child is created.
All other nodes are internal nodes.

Put $h:=\alpha2^k kd$.
For an instance $\widehat I$ at an internal node, define
\[
 p(\widehat I)=
 \sum_{i\in[\alpha]}
 \sum_{\sigma\in\{0,1\}^k}
 \sum_{j\in[k]}
 \left|
 \left\{
 E\in\phi_i\setminus\psi_{i,\sigma}:
 P_j\subseteq\var(E)
 \right\}
 \right|.
\]
At every internal node, all candidate sets have size greater than $d$.
For fixed $i,\sigma,j$, at most $d$ clauses of $\phi_i$ can contain
all variables of $P_j$.
Otherwise, since $|P_j|\ge d+1$, the graph $\inc(\phi_i)$ contains
a $K_{d+1,d+1}$, contradicting $d$-degeneracy.
There are $\alpha2^k k$ triples $(i,\sigma,j)$, so
$0\le p(\widehat I)\le h$.

Every edge between two internal nodes corresponds
to the fourth branching rule, since the first
three rules create singleton candidate sets.
Along such an edge, previously counted clauses
remain counted because candidate sets only shrink.
The newly deleted clause contributes an additional
term for the selected label and candidate-set index.
Thus $p$ increases by at least one.

Since $0\leq p\leq h$ at every internal node,
a root-to-leaf path contains at most $h$ edges
between internal nodes and at most one final
edge to a leaf.
Its length is therefore at most $h+1$.

\subparagraph{Running time and output guarantees.}
Each internal node has at most $5k$ children, and each leaf
outputs at most $d$ instances.
Thus the number of nodes and output instances is at most
$d(5k)^{h+2} = 2^{\alpha d\log d\,2^{O(k)}}$.
There are $\alpha2^k$ formula entries, each a subformula
of an original formula, and candidate sets only shrink.
Hence every recursive and output instance has size at most
$2^{O(k)}|I_0|$.
All incidence graphs remain $d$-degenerate because they
are obtained by deleting clause vertices from the original
incidence graphs.

Using the $O(d^4(n+m))$-time algorithm of Lemma~\ref{lem:find-k22} for finding a
$K_{2,2}$, processing each node, including the construction of
its children or outputs, takes
$d^{O(1)}2^{O(k)}|I_0|$ time.
Multiplying by the number of nodes gives the claimed total running time.

Every output instance retains the original variable set $V$, has $k$ candidate sets and at most $\alpha2^k$ formulas.
By the base cases, either all its incidence graphs are $K_{2,2}$-free,
or it has a specified singleton candidate set $\{v_s\}$.
In the latter case, every solution contains $v_s$.
Together with the correctness argument, this proves all the
required properties.
\end{proof}

In the remaining section we prove \cref{thm:bounded-degeneracy} by combining \cref{lem:four-cycle-preprocessing} with
\cref{lem:colorful-linear}.

\coloredboundeddegfpt*

\begin{proof}[Proof of \cref{thm:bounded-degeneracy}]
If $d=1$, every input incidence graph is a forest,
so the instance is a yes-instance if and only if
every candidate set is nonempty.
This takes $O(|I|)$ time.
Hence assume $d\geq2$.

We give a recursive algorithm.
Let
$I'=(\Psi,V,Q_1,\ldots,Q_{k'})$ where $    \Psi=(\psi_1,\ldots,\psi_{\alpha'})$,
be the instance at a recursive call.
If some $Q_j$ is empty, return no.
If $k'=0$, return yes if and only if every
$\inc(\psi_i)$ is a forest.

Otherwise, apply \cref{lem:four-cycle-preprocessing}
to $I'$.
If it decides $I'$, return its answer.
Otherwise, write its output family as
$ I_s=(\Psi_s,V,R^s_1,\ldots,R^s_{k'})$ for $s\in[\ell]$.
Process each $I_s$ as follows.
\begin{enumerate}
    \item If every formula in $\Psi_s$ has a
    $K_{2,2}$-free incidence graph, solve $I_s$
    using \cref{lem:colorful-linear}.

    \item Otherwise, by item~(iii) of
\cref{lem:four-cycle-preprocessing}, the output
instance $I_s$ has a singleton candidate set.
Choose an index $j$ with $R^s_j=\{v\}$.
    Put
    \[
        \Psi^v=
        \bigl(
          \psi_1[v=0],\psi_1[v=1],
          \ldots,
          \psi_{\alpha'}[v=0],\psi_{\alpha'}[v=1]
        \bigr),
    \]
    and recursively solve
    \[
        I_s^-=
        \bigl(
          \Psi^v,V,
          R^s_1,\ldots,R^s_{j-1},
          R^s_{j+1},\ldots,R^s_{k'}
        \bigr).
    \]
    The reducts are taken from the input tuple
    $\Psi$, not the output tuple $\Psi_s$.
    All other candidate sets remain unchanged.

For each $b\in\{0,1\}$, scan the formulas
$\psi_1,\ldots,\psi_{\alpha'}$ to construct their
semi-reducts under $v=b$. To merge identical clauses,
initially put all surviving clause occurrences of each
formula into one group.
For each literal $\ell$, build its \emph{occurrence list}
by storing a pointer to every surviving clause occurrence
that contains $\ell$, across all formulas.

Process the literals one at a time. For each literal
$\ell$, scan its occurrence list. In each group reached
by this scan, move the clauses containing $\ell$ into
a new group. The other clauses stay in their current
group. With linked lists for the groups and a pointer
from each clause to its current group, this takes
constant time per literal occurrence.

At the end, two clauses of the same formula belong to
the same group exactly when they contain the same
literals. Retain one clause from each nonempty group.
Initializing the occurrence lists and groups takes
$O(|V|+\alpha')$ time, and the remaining work is linear
in the total size of the input formulas. Thus constructing
both reducts of all formulas takes
$O(|V|+\alpha'+\sum_{i=1}^{\alpha'}|\psi_i|)$ time.

\end{enumerate}
Return yes if and only if at least one of these
computations returns yes.

\paragraph{Correctness.}
The base cases are immediate.
By \cref{lem:four-cycle-preprocessing}\,(i),
$I'$ is a yes-instance if and only if some $I_s$
is a yes-instance.
The first case is handled correctly by
\cref{lem:colorful-linear}.

For the second case, item~(iv) of
\cref{lem:four-cycle-preprocessing} permits replacing
$\Psi_s$ by $\Psi$ without changing the validity
of any choice from the candidate sets of $I_s$.
Every such choice selects $v$ from $R^s_j$.
Let $Y$ be the set of variables chosen from the
remaining candidate sets.
Then $\{v\}\cup Y$ is a strong forest backdoor
of every formula in $\Psi$ if and only if $Y$
is a strong forest backdoor of every formula
in $\Psi^v$, since both assignments to $v$
are considered.
This remains valid when $v\in Y$, because $v$
is absent from every formula in $\Psi^v$.
Thus $I_s$ and $I_s^-$ are equivalent.
Each recursive call removes one candidate set,
so correctness follows by induction on $k'$.

\paragraph{Running time.}
Taking reducts preserves non-tautologicality
and $d$-degeneracy.
Each recursive step removes one candidate set
and doubles the number of formulas.
Consequently, at depth $t$,
$ k'=k-t$, $\alpha'\leq\alpha2^t$ and $ |I'|\leq2^{O(t)}|I|$.
The recursion therefore has depth at most $k$.

At each call, preprocessing takes
$ 2^{\alpha'd\log d\,2^{O(k')}}|I'|
    \leq
    2^{\alpha d\log d\,2^{O(k)}}|I|$
    time and produces at most
$2^{\alpha d\log d\,2^{O(k)}}$ output instances.
Each output has at most
$\alpha'2^{k'}\leq\alpha2^k$
formulas and size at most
$2^{O(k')}|I'|\leq2^{O(k)}|I|$.
Hence solving any $K_{2,2}$-free output using
\cref{lem:colorful-linear} takes
$2^{\alpha2^{O(k)}}|I|$ time.

There is therefore a constant $C$ such that, with
$ B:=2^{\alpha d\log d\,2^{Ck}}$,
every call creates at most $B$ recursive children
and performs at most $B|I|$ work, including
preprocessing, constructing children, and solving
all its $K_{2,2}$-free outputs.
The recursion tree has at most $B^{k+1}$ nodes.
Thus the total running time is at most
$ B^{k+2}|I|
    =2^{\alpha d\log d\,2^{O(k)}}|I|$,
as required.

\paragraph{Returning a backdoor.}
The algorithms also return a witnessing tuple
whenever they report yes.
Retain the successful choices at enumeration leaves
and insert each fixed variable into its removed
candidate-set position when returning from recursion.
Choices obtained from a preprocessing output remain
valid for its input by item~(iv) of
\cref{lem:four-cycle-preprocessing}.
These operations preserve the stated running time.
\end{proof}

\section{The tightness of the double-exponential factor}\label{sec:eth}

In this section, we prove \cref{thm:eth}.

\strongforesteth*

Recall that a Boolean assignment satisfies a CNF formula if every
clause contains a literal that is true under the assignment.
A $3$-CNF formula has at most three literals per clause.

We begin by defining the {\sc $3$-SAT} and {\sc Set Cover} problems which are used to give the reduction that proves \cref{thm:eth}.

\defproblem{\textsc{$3$-SAT}}{A $3$-CNF formula $\psi$}{Does $\psi$ admit a satisfying Boolean assignment?} 

\defparproblem{\textsc{Set Cover}}
  {A universe $U$, a family
   $\mathcal{F}$ of subsets of $U$, and a positive
   integer $K$.}
  {$K$}
  {Does there exist $\mathcal{F}' \subseteq \mathcal{F}$ with $|\mathcal{F}'|\leq K$ such that
   $\bigcup_{F\in \mathcal{F}'}F = U$?}

  Without loss of generality assume that  $|U| \geq 2$, every set in $\mathcal{F}$ is nonempty, every
    element of $U$ belongs to some set in $\mathcal{F}$, and
    $1\leq K\leq  |\mathcal{F}|$.
    In this section we prove \cref{lem:eth-set-cover}.

The proof of \cref{thm:eth} composes two reductions.
First, we transform a \textsc{Set Cover} instance
$(U,\mathcal{F},K)$ into an equivalent instance $(\phi,k)$
of \strongback{}, where
$k=K+\lceil\log|U|\rceil$.
Second, we group the variables of an $N$-variable
$3$-CNF formula into $r$ blocks to obtain a
\textsc{Set Cover} instance with budget $K=r$,
a universe of polynomial size, and explicit encoding size
$2^{O(N/r+\log N)}$.
For each fixed positive integer $c$, choosing
$r=c\lceil\log N\rceil$ gives a backdoor budget
$O(c\log N)$.
We show that a $2^{2^{o(k)}}|\phi|^{O(1)}$-time algorithm
for \strongback{}, combined with these reductions and
a dovetailing argument~\cite{DBLP:books/daglib/0086373}
over the choices of $c$, would
solve $3$-SAT in $2^{o(N/\log N)}$ time.

\begin{lemma}[Reduction from Set Cover]\label{lem:eth-set-cover}
    Given a \textsc{Set Cover} instance $(U,\mathcal{F},K)$,
    one can construct in polynomial time a non-tautological
    $3$-CNF formula $\phi$ and an integer
    $k=K+\lceil\log |U|\rceil$
    such that $(U,\mathcal{F},K)$ is a yes-instance of {\sc Set Cover} if and only if $(\phi,k)$ is a yes-instance of \strongback. 
    \end{lemma}
    \begin{proof}

    Let $(U,\mathcal{F},K)$ be a \textsc{Set Cover} instance.
    Without loss of generality, let $U=\{0,1, \ldots, n-1\}$.
    Set $d:= \lceil \log n \rceil$.

For each $F\in\mathcal{F}$, introduce a Boolean variable $x_F$.
Choosing $x_F$ as a backdoor variable corresponds to choosing
$F$ for the set cover.
Also introduce selector variables $S:=\{s_1,\ldots,s_d\}$.
    For each $e\in U$, 
    let
$\operatorname{code}(e)\in\{0,1\}^d$
be the unique $d$-bit binary representation of $e$.
For any $i \in [d]$, $\operatorname{code}(e)_i$ denotes the $i$-th position of the bit-string $\operatorname{code}(e)$.
For every $i \in [d]$, define literal
$$
  \ell_{e,i}:=
  \begin{cases}
    s_i,      & \text{if }\operatorname{code}(e)_i=0,\\
    \neg s_i, & \text{if }\operatorname{code}(e)_i=1.
  \end{cases}
$$
Thus all literals $\ell_{e,1},\ldots,\ell_{e,d}$ are false
under an assignment $\sigma:S\to\{0,1\}$ if and only if
$\sigma(s_i)=\operatorname{code}(e)_i$ for every $i\in[d]$.

\paragraph{Element cycles.}
For each $e\in U$, let
$ T_e:=\{x_F:F\in\mathcal{F},\ e\in F\}$.
Introduce a fresh Boolean variable $a_e$.
Construct a simple cycle $Z_e$ whose distinguished vertices
are $a_e$ and the variables in $T_e$, in any fixed order.
Between every two successive distinguished vertices,
place $d+1$ fresh subdivision vertices, so that they are
separated by $d+2$ edges.
Every subdivision vertex is also a Boolean variable.
The variable $a_e$ and all subdivision variables are private
to $e$; only the variables $x_F$ are shared between element
cycles.
Thus $Z_e$ has exactly $(|T_e|+1)(d+2)$ vertices and
the same number of edges.
Since $T_e\neq\emptyset$ and $d\geq1$, the construction
produces a simple cycle even when $|T_e|=1$.

Choose a direction around $Z_e$.
Between each pair of successive distinguished vertices,
label the $d+2$ edges, in that direction, by
$ 1,2,\ldots,d,0,0$.
For every edge $uv$ of $Z_e$, add the clause
$$
  C_{e,uv}:=
  \begin{cases}
    \{u,v,\ell_{e,i}\},
      & \text{if the edge has label }i\in[d],\\
    \{u,v\},
      & \text{if the edge has label }0.
  \end{cases}
$$

Let $\sigma:S\to\{0,1\}$ be the assignment defined by
$\sigma(s_i)=\operatorname{code}(e)_i$ for every $i\in[d]$.
Under this assignment, every literal $\ell_{e,i}$ is false.
For every edge $uv\in E(Z_e)$, the clause $C_{e,uv}$
reduces to $\{u,v\}$ under $\sigma$.
Indeed, a binary clause remains unchanged, while a
ternary clause loses its false selector literal.
Hence the incidence graph formed by these residual
clauses is a subdivision of $Z_e$.
If an assignment $\sigma:S\to\{0,1\}$ differs from
$\operatorname{code}(e)$ at some position $i\in[d]$,
then $\ell_{e,i}$ is true under $\sigma$.
Every segment of $Z_e$ between successive distinguished
vertices contains an edge labelled $i$.
The clause corresponding to that edge is therefore
satisfied and deleted, breaking every such segment.

\paragraph{Selector-forcing gadgets.}
For every $i\in[d]$ and $j\in[k+1]$, introduce two fresh
private variables $p_{i,j},q_{i,j}$ and add the clauses
$\{s_i,p_{i,j}\},
  \{s_i,q_{i,j}\},
  \{p_{i,j},q_{i,j}\}$.

Let $\phi$ consist of all clauses constructed above.
Observe that every clause is non-tautological and has size at most three.
Also 
$ |\phi|
  =O\bigl((d+1)(n+\sum_{F\in\mathcal{F}}|F|)+d(k+1)\bigr) = O(|U||\mathcal{F}| \log |U|)$.
Therefore the construction has polynomial size and
can be carried out in polynomial time.

\begin{claim}\label{claim:backdoor}
     Every strong forest backdoor $X$ of
$\phi$ with $|X|\leq k$ contains $S$.
\end{claim}
\begin{proof}
Recall that $S= \{s_1, \ldots, s_d\}$.
For the sake of contradiction suppose that $s_i\not \in X$ for some $i \in [d]$.
Since the pairs $\{p_{i,j},q_{i,j}\}$, $j\in[k+1]$,
are pairwise disjoint and $|X|\le k$, there exists
$j\in[k+1]$ such that
$ X\cap\{p_{i,j},q_{i,j}\}=\emptyset$.

Since $s_i,p_{i,j},q_{i,j} \not \in X$, the three clauses
$ \{s_i,p_{i,j}\},
  \{s_i,q_{i,j}\},
  \{p_{i,j},q_{i,j}\}
$ (corresponding to the selector-forcing gadget)
survive unchanged under every assignment to $X$.
Their incidence graph is a cycle of length six,
contradicting that $X$ is a strong forest backdoor.

\end{proof}

\begin{claim}\label{claim:remaining-cycle}
Let $\sigma:S\to\{0,1\}$ be an assignment.
Suppose that, for some $e\in U$,
$\sigma(s_i)=\operatorname{code}(e)_i$ for every $i\in[d]$.
Then, for every $Y\subseteq\var(\phi)\setminus S$ and every
assignment $\tau:Y\to\{0,1\}$,
$\inc(\phi[\sigma\cup\tau])$ is a forest if and only if $  Y\cap V(Z_e)\neq \emptyset$.

If no such element $e$ exists, then
$\inc(\phi[\sigma\cup\tau])$ is a forest for every such
$Y$ and $\tau$.
\end{claim}
\begin{proof}
    Fix an assignment $\sigma:S\to\{0,1\}$.
Observe that every non-selector variable occurs only positively
in $\phi$, and every clause of size three contains a selector
literal. Hence every clause surviving in $\phi[\sigma]$
is positive and has size at most two.

Let $H_\sigma$ be the graph whose vertices are the
non-selector variables and for every binary clause
$\{u,v\}$ of $\phi[\sigma]$, add the edge $uv$ to $H_\sigma$.

We show that $H_\sigma$ is a forest if $\sigma$ matches
no element code, and otherwise has exactly one cycle,
namely $Z_e$ for the unique element $e$ whose code
matches $\sigma$.

Suppose that there exists $e\in U$ such that
$\sigma(s_i)=\operatorname{code}(e)_i$ for every $i\in[d]$.
Since the element codes are distinct, such an element is unique.
Every literal $\ell_{e,i}$ is false under $\sigma$.
Thus, for every $uv\in E(Z_e)$, the clause $C_{e,uv}$
reduces to $\{u,v\}$, and $Z_e$ is a subgraph of $H_\sigma$.

Consider any element $f\in U\setminus\{e\}$.
There exists $i\in[d]$ such that $\sigma(s_i)\ne\operatorname{code}(f)_i$.
By the definition of $\ell_{f,i}$, this literal is true under $\sigma$.
On every subpath of $Z_f$ between successive distinguished
vertices, there is an edge labelled $i$.
The clause corresponding to this edge is satisfied and deleted.
Consequently, the surviving edges of $Z_f$ form a forest
in which each component contains at most one distinguished
vertex, and hence at most one shared variable $x_F$.

By construction, $a_e$ and the subdivision variables of $Z_e$
occur in no other element cycle. Hence, for distinct $e,f\in U$,
$ V(Z_e)\cap V(Z_f)
  =\{x_F:F\in\mathcal{F},\ e,f\in F\}$.
As shown above, each component formed by the surviving edges
of a cycle $Z_f$ with $f\ne e$ contains at most one variable
of the form $x_F$.
Joining these components at shared variables therefore produces
a forest, each component of which meets $Z_e$ in at most one
vertex.
Consequently, adding these components to $Z_e$ creates no
additional cycle.
It follows that the element clauses contribute exactly
one cycle to $H_\sigma$, namely $Z_e$.
If no element has code matching $\sigma$, the same argument
applies to every element cycle, and their surviving edges
form a forest.

For every $i\in[d]$ and $j\in[k+1]$, the clause
$\{p_{i,j},q_{i,j}\}$ survives unchanged in $\phi[\sigma]$,
since neither of its variables belongs to $S$.
The other two clauses are either deleted or become unit
clauses, depending on $\sigma(s_i)$.
Since these private variables occur nowhere else, these
clauses contribute only separate tree components. 
Thus $H_\sigma$ is a forest when $\sigma$ matches no
element code; otherwise, its unique cycle is $Z_e$.

Now fix $Y\subseteq\var(\phi)\setminus S$ and an assignment
$\tau:Y\to\{0,1\}$.
A binary clause $\{u,v\}$ of $\phi[\sigma]$ remains binary
after applying $\tau$ if and only if $u,v\notin Y$.
Every other surviving clause is unit or empty.
Hence the binary clauses of $\phi[\sigma\cup\tau]$
correspond exactly to the edges of $H_\sigma-Y$.
For each edge $uv$ of $H_\sigma-Y$, the corresponding
binary clause $C=\{u,v\}$ gives the path $u,C,v$ in the
incidence graph. Thus each edge is replaced by a path
of length two, which preserves the existence of cycles.
Unit clauses contribute only leaves to the incidence graph,
and empty clauses contribute isolated vertices.
Therefore,
$\inc(\phi[\sigma\cup\tau])$
is a forest
 if and only if $H_\sigma-Y$  is a forest.
If $\sigma$ matches the code of $e$,that is
if $\sigma(s_i)=\operatorname{code}(e)_i$ for every $i\in[d]$, the graph $H_\sigma$
has the unique cycle $Z_e$, so this condition holds
precisely when $Y\cap V(Z_e)\ne\emptyset$.
If $\sigma$ matches no element code, $H_\sigma$ is already
a forest, and the condition holds for every $Y$ and $\tau$.
\end{proof}

\paragraph{Correctness.}
For the forward direction, let
$\mathcal{F}'\subseteq\mathcal{F}$ be a set cover of $(U, \mathcal{F})$ of size at most $K$.
Set
$X:=S\cup\{x_F:F\in\mathcal{F}'\}$.
Then $|X| \leq d+K = k$.
We claim that $X$ is a strong forest backdoor for the formula $\phi$.
Fix any assignment $\alpha$ to $X$, and let
$\sigma:=\alpha|_S$.
If there is no element $e\in U$ such that
$\sigma(s_i)=\operatorname{code}(e)_i$ for every $i\in[d]$,
then \cref{claim:remaining-cycle} implies that $\inc(\phi[\alpha])$ is a forest.
Otherwise, there exists a unique $e\in U$ such that
$\sigma(s_i)=\operatorname{code}(e)_i$ for every $i\in[d]$.
Since $\mathcal{F}'$ covers $U$, some $F\in\mathcal{F}'$
contains $e$.
By construction, $x_F\in X\setminus S$ and $x_F\in V(Z_e)$.
Hence $(X\setminus S)\cap V(Z_e)\ne\emptyset$.
Applying \cref{claim:remaining-cycle} with
$Y:=X\setminus S$ and $\tau:=\alpha|_Y$ shows that
$\inc(\phi[\alpha])$ is a forest.
Since $\alpha$ was an arbitrary assignment to $X$,
the set $X$ is a strong forest backdoor of size at most $k$.

For the backward direction,
 let $X$ be a strong forest
backdoor of $\phi$ of size at most $k$.
From \cref{claim:backdoor}, $S\subseteq X$.
Put $Y:=X\setminus S$, and note that $|Y|\le K$.
We claim that, for every $e\in U$, the set $Y$ contains
at least one variable corresponding to a vertex of the cycle $Z_e$.
Fix $e\in U$ and let $\sigma:S\to\{0,1\}$ be defined by
$\sigma(s_i)=\operatorname{code}(e)_i$ for every $i\in[d]$.
Choose any assignment $\tau:Y\to\{0,1\}$.
Since $X=S\cup Y$ is a strong forest backdoor,
$\inc(\phi[\sigma\cup\tau])$ is a forest.
\cref{claim:remaining-cycle} therefore implies that
$Y\cap V(Z_e)\ne\emptyset$.
Thus $Y$ contains at least one variable corresponding
to a vertex of $Z_e$.
Since $e\in U$ was arbitrary, this holds for every element.

We now construct a family $\mathcal{F}'\subseteq\mathcal{F}$
from $Y$.
For every variable $x_F\in Y$, include $F$ in $\mathcal{F}'$.
For every variable in $Y$ that is private to an element
cycle $Z_e$, include an arbitrary set of $\mathcal{F}$
containing $e$.
Such a set exists because every element of $U$ belongs
to at least one set in $\mathcal{F}$.
Each variable contributes at most one set, so, after
discarding repetitions,
$|\mathcal{F}'|\le |Y|\le K$.

To see that $\mathcal{F}'$ covers $U$, fix $e\in U$.
As proved above, $Y$ contains a variable corresponding
to a vertex of $Z_e$.
If this variable is $x_F$, then $e\in F$ by construction,
and $F\in\mathcal{F}'$.
Otherwise, the variable is private to $Z_e$, and the
construction of $\mathcal{F}'$ includes a set containing $e$.
Thus every element of $U$ is covered by $\mathcal{F}'$.
Hence $(U,\mathcal{F},K)$ is a yes-instance of
\textsc{Set Cover}.
\end{proof}

\begin{proof}[Proof of \cref{thm:eth}]
We give a reduction from {\sc $3$-SAT} to \strongback.
Let $\psi$ be a $3$-CNF formula on $n$ variables.
Without loss of generality, we can assume that $\psi$ has no tautological clauses and all clauses of $\psi$ are distinct. If not, then tautological and repeated clauses can be deleted without loss of generality for solving the {\sc $3$-SAT} problem. 
Thus we may assume that $\psi$ has $O(n^3)$ clauses.
Set $L:=\lceil\log n\rceil$.
We handle formulas containing an empty clause
and the finitely many small values of $n$ directly.
Hence assume that every clause is nonempty and
that $\lceil\log n\rceil^2\leq n$.

For any integer $r$ with $L\leq r\leq L^2$, partition
the variables into $r$ nonempty blocks
$B_1,\ldots,B_r$, each of size at most $\lceil n/r\rceil$.
Since $r\leq L^2$, we have $\log n=O(n/r)$.

Construct a \textsc{Set Cover} instance $(U,\mathcal{F},K)$
as follows. 
The universe $U$ contains a distinct element $b_i$ for each
block $B_i$, $i \in [r]$,
and an element $c_C$ for each clause $C$ of $\psi$.
That is $U = \{b_i : i \in [r]\} \cup \{c_C: C \text{ is a clause of } \psi\}$.

For every $i\in[r]$ and assignment $\theta : B_i \to \{0,1\}$, 
introduce
the set
$$
  F_{i,\theta}
  =\{b_i\}\cup
    \{c_C:\theta\text{ satisfies a literal of }C
                    \text{ whose variable belongs to }B_i\}.
$$
Let $\mathcal{F}$ consist of these sets $F_{i, \theta}$,
and set $K:=r$.
Note that $|U| = O(n^3)$ and $|\mathcal{F}| \leq  r2^{\lceil n/r\rceil}$.
The instance can therefore be constructed in time $2^{O(n/r)}$.

We now claim that $\psi$ is satisfiable if and only if
$(U,\mathcal{F},K)$ is a yes-instance of {\sc Set Cover}.
Let $\sigma$ be a satisfying assignment for $\psi$.
For each $i \in [r]$, 
let $\sigma_{B_i}$ be the assignment $\sigma$ restricted to the variables of the block $B_i$.
Then observe that $\mathcal{F'} := \{F_{i, \sigma_{B_i}} : i \in [r]\}$ is a set cover of 
the instance $(U,\mathcal{F})$ of size $r = K$.

For the backward direction let $\mathcal{F}' \subseteq \mathcal{F}$
be a set cover of $(U,\mathcal{F})$ of size $K=r$.
Since only sets associated with block $B_i$ contain $b_i$,
covering all $r$ block elements with at most $r$ sets
requires $\mathcal{F}'$ to contain exactly one set
associated with each block.
Let $\mathcal{F}' = \{F_{1, \theta_1}, \ldots,  F_{r, \theta_r}\}$.
The assignments $\theta_1,\ldots,\theta_r$ have pairwise
disjoint domains $B_1,\ldots,B_r$ and therefore combine
into an assignment $\sigma$ to all variables of $\psi$.
This assignment $\sigma$ satisfies all clauses of $\psi$ because every clause element is covered by $\mathcal{F}'$.

From \cref{lem:eth-set-cover}, given this {\sc Set Cover} instance $(U,\mathcal{F},K)$, we can construct, in time polynomial in the instance $(U,\mathcal{F},K)$, an instance $(\phi,k)$ of \strongback, 
where $\phi$ is a non-tautological $3$-CNF
formula and
$ k=K+\lceil\log |U|\rceil
    =r+O(\log n)$.
    
Combining both reductions, we conclude that given an instance
$\psi$ of {\sc $3$-SAT} on $n$ variables, we can construct
an instance $(\phi,k)$ of \strongback{} in time $2^{O(n/r)}$,
of size $2^{O(n/r)}$, where $k=r+O(\log n)$,
such that $\psi$ is satisfiable if and only if $(\phi,k)$
is a yes-instance of \strongback{}.
Moreover, $\phi$ is non-tautological.

Suppose that \strongback{} admits an algorithm running in
time $2^{2^{h(k)}}|\phi|^{O(1)}$, where $h(k)=o(k)$.
For each integer $c\in[L]$, apply the preceding construction
with $r=cL$ and run the assumed algorithm on the resulting
instance.
Fix any positive integer $c$.
For each $n$, we run computations with block counts
$r=L,2L,\ldots,L^2$.
For any fixed positive integer $c$, the block count $r=cL$
appears in this list whenever $L\geq c$, which holds for
all sufficiently large $n$.
The constructed instance has
$k=cL+O(\log n)=O(c\log n)$.
Since $c$ is fixed and $h(k)=o(k)$, we obtain
\[
    2^{h(k)}=n^{o(1)}=o(n/\log n).
\]
Including the construction time, the running time of this
computation is at most
\[
    2^{2^{h(k)}+O(n/r)}
    \leq
    2^{(C/c+o(1))\,n/\log n},
\]
where $C$ is a constant independent of $c$.

We dovetail (see for example~\cite{DBLP:books/daglib/0086373})
the computations for $c=1,\ldots,L$,
each consisting of constructing its instance $(\phi,k)$
and running the assumed algorithm on that instance.
In each round, perform one step of each computation,
and stop when the first computation finishes\footnote{Dovetailing interleaves the computations,
performing one step of each per round and stopping
when the first finishes.
For a standard example of this simulation, see
\cite[Proof of Theorem~4.22]{DBLP:books/daglib/0086373}.
Here we use it to obtain a single algorithm
without choosing $c$ in advance.
Every fixed $c$ is eventually included among
the simulated computations, allowing the constant
$C/c$ in the exponent to become arbitrarily small.
Interleaving the $L$ computations incurs only
polynomial overhead.}.
Every computation correctly decides whether $\psi$ is
satisfiable.
The interleaving increases the running time by only a
polynomial factor in $n$.

For every $\varepsilon>0$, choose a fixed positive integer
$c$ such that $C/c<\varepsilon/2$.
For all sufficiently large $n$, this computation is included
and the total running time is at most
$2^{\varepsilon n/\log n}$.
Thus the resulting single algorithm solves $3$-SAT in
$2^{o(n/\log n)}$ time, contradicting the assumption of
the theorem.
\end{proof}

\section{Extension to treewidth $2$ or more is impossible under \FPT$\neq$\Wone}\label{sec:fptneqwone}

In this section, we prove \cref{thm:hardness}.

\strongtreewidthhardness*

To prove \cref{thm:hardness} we first prove that {\sc Simultaneous Treewidth-$t$ Deletion}, which is defined below, is \Wone-hard.

\defparproblem{\textsc{Simultaneous Treewidth-$t$ Deletion}}
  {Graphs $G_1,\ldots,G_{\alpha}$ on the same vertex set $V$,
   and a nonnegative integer $K$.}
  {$K+\alpha$}
  {Does there exist $D\subseteq V$ with $|D|\le K$ such that
   $\operatorname{tw}(G_a-D)\le t$ for every $a\in[\alpha]$?}

{\sc Simultaneous Treewidth-$t$ Deletion} was shown to be FPT when $t=1$ in~\cite{DBLP:journals/toct/AgrawalLMS18}. Here we show that the problem becomes \Wone-hard for any $t \geq 2$.
This is then used as a source problem to prove the \Wone-hardness result in \cref{thm:hardness}.

\lemmasimtw*

To prove \cref{lem:simultaneous-treewidth-hardness} we give a parameterized reduction from {\sc Paired Cut}, defined below.

\defparproblem{Paired Cut}
  {Two undirected graphs $H_1,H_2$, distinct terminals
   $s_i,t_i\in V(H_i)$ for each $i\in\{1,2\}$, a family
   $\mathcal B\subseteq E(H_1)\times E(H_2)$ of edge pairs
   in which each edge occurs in at most one pair, and a
   nonnegative integer $\ell$.}
  {$\ell$}
  {Does there exist $\mathcal B'\subseteq\mathcal B$ with
   $|\mathcal B'|\le \ell$ such that, for each $i\in\{1,2\}$,
   deleting the edges of $H_i$ belonging to pairs in
   $\mathcal B'$ separates $s_i$ from $t_i$?}

Without loss of generality assume that  $1 \leq \ell \leq |\mathcal B|$.
Dabrowski, Jonsson, Ordyniak, Osipov, Pilipczuk and Sharma~\cite[Lemma~17]{DBLP:conf/iwpec/DabrowskiJOOPS23} showed that {\sc Paired Cut} is \Wone-hard. We observe that their hardness reduction actually shows hardness even when both $H_1$ and $H_2$ are simple graphs and have treewidth two. 
This is captured in \cref{lem:paired-cut-series-parallel}.

\begin{definition}[Series-parallel graphs]
A \emph{two-terminal graph} is a graph with two distinct
specified vertices, called its terminals.
A two-terminal graph is \emph{series-parallel} if it can
be obtained from a single edge, whose endpoints are the
terminals, by repeatedly applying the following operations
to vertex-disjoint two-terminal graphs with terminal pairs
$(u_1,v_1)$ and $(u_2,v_2)$:
\begin{enumerate}
  \item \emph{Series composition:}
  identify $v_1$ with $u_2$, and take $u_1,v_2$ as the
  terminals of the resulting graph.

  \item \emph{Parallel composition:}
  identify $u_1$ with $u_2$ and $v_1$ with $v_2$, and take
  the two identified vertices as the terminals.
\end{enumerate}
Parallel edges are allowed in this definition.
\end{definition}

Every series-parallel graph has treewidth at most
two~\cite[Section~2.2]{DBLP:journals/cj/BodlaenderK08}.

\begin{lemma}[\cite{DBLP:conf/iwpec/DabrowskiJOOPS23}]\label{lem:paired-cut-series-parallel}
\textsc{Paired Cut} is \Wone-hard parameterized by $\ell$,
even when both input graphs are simple (that is without parallel edges) and series-parallel.
In particular, the hardness holds when both input graphs
have treewidth two.
\end{lemma}
\begin{proof}[Proof Sketch]
Dabrowski et al.~\cite[Lemma~17]{DBLP:conf/iwpec/DabrowskiJOOPS23}
prove \Wone-hardness of \textsc{Paired Cut}.
We observe that the graphs produced by their reduction
are series-parallel.

The construction in Section~6.1
of~\cite{DBLP:conf/iwpec/DabrowskiJOOPS23} first defines
two directed graphs, whose underlying undirected graphs form the input to
\textsc{Paired Cut}.
Each directed graph is a union of subgraphs
that share only the two terminals.
Within each such subgraph, the construction places internally
disjoint paths between consecutive vertices.
The union of paths with the same endpoints is a parallel
composition. Joining these subgraphs at consecutive vertices
is a series composition.
Finally, these subgraphs share only the two
terminals, so their union is a parallel composition.
Since every path is a series composition of edges, both
constructed graphs are series-parallel
and therefore have treewidth at most two.

Our subsequent reduction uses simple input graphs.
However, the construction above may produce
parallel edges, which cannot simply be merged:
they represent distinct edges in the pairing,
and merging them could change the feasible cuts.
We therefore subdivide every edge once, using
a fresh vertex for each edge.

For every original paired edge, designate one
of its two replacement edges.
Replace each original pair by the pair of
designated replacement edges, leaving the other
replacement edges unpaired.
Selecting a pair therefore breaks exactly the
two paths corresponding to its original edges.
Consequently, the feasible paired cuts and the
budget are preserved.
The resulting graphs are simple and remain
two-terminal series-parallel.

Composing their reduction with this subdivision
step establishes \Wone-hardness even when both
input graphs are simple and two-terminal
series-parallel.
\end{proof}

\begin{proof}[Proof of \cref{lem:simultaneous-treewidth-hardness}]
To prove the lemma, 
we give a parameter-preserving reduction from {\sc Paired Cut} to {\sc Simultaneous Treewidth-$t$ Deletion}.
From \cref{lem:paired-cut-series-parallel} we can assume that the input graphs in the {\sc Paired Cut} instance both have treewidth two.

\paragraph{Construction.}
Let $(H_1, H_2, s_1, t_1, s_2, t_2, \mathcal{B}, \ell)$ be an instance of {\sc Paired Cut}, where both $H_1$ and $H_2$ are series-parallel graphs.

We reduce from the restriction of \textsc{Paired Cut}
in which both input graphs are two-terminal simple series-parallel
networks with their designated terminals.

For every pair $b\in\mathcal B$, introduce a vertex $x_b$,
which we call a \emph{choice vertex}.
For every $i\in\{1,2\}$ and $j\in[\ell+1]$, let
$ V_{i,j}:=\{v_{i,j}:v\in V(H_i)\}$ and 
 let $  W_{i,j}:=\{w_{i,j,1},\ldots,w_{i,j,t}\}$,
be sets of fresh vertices, all distinct from one another
and from the choice vertices. (All these sets and the choice vertices
are pairwise disjoint.)
Define the common vertex set
$$
  V:=\{x_b:b\in\mathcal B\}
     \uplus\biguplus_{i=1}^{2}\biguplus_{j=1}^{\ell+1}
       (V_{i,j}\uplus W_{i,j}).
$$
For each $i\in\{1,2\}$ and $j\in[\ell+1]$, construct a graph
$G_{i,j}$ on $V$ as follows. First, place a copy of $H_i$ on
$V_{i,j}$,
that is
for every edge $uv\in E(H_i)$, add the edge
$u_{i,j}v_{i,j}$ in $G_{i,j}$. Then, for each pair $b=(e_1,e_2)\in\mathcal B$, 
replace
the copy of $e_i=uv$ by the path
$ u_{i,j},\ x_b,\ v_{i,j}$.
Denote by $\widetilde H_i^{(j)}$ the graph obtained from
the copy of $H_i$ on $V_{i,j}$ after these subdivisions.
Its terminals are $s_{i,j}$ and $t_{i,j}$, the copies of
$s_i$ and $t_i$, respectively.
Next, make $W_{i,j}$ a clique and join every vertex of
$W_{i,j}$ to both terminals.
Thus the added edges form a copy of
$K_{t+2}-s_{i,j}t_{i,j}$ on
$W_{i,j}\cup\{s_{i,j},t_{i,j}\}$.
All vertices outside this network and $W_{i,j}$ are isolated
in $G_{i,j}$.

Finally, set $K:=\ell$. The output consists of the
$\alpha:=2(\ell+1)$ graphs $G_{i,j}$ on common vertex set $V$,
with deletion budget $K$.
For each $(i,j)$, every vertex of $V_{i,j}\cup W_{i,j}$
is isolated in all graphs other than $G_{i,j}$.
The choice vertices $x_b$, however, occur in the subdivided
network of every graph.

\begin{claim}\label{claim:network-treewidth}
For every $i\in\{1,2\}$, $j\in[\ell+1]$, and
$Z\subseteq\{x_b:b\in\mathcal B\}$,
$\operatorname{tw}(G_{i,j}-Z)\le t$ if and only if
$\widetilde H_i^{(j)}-Z$ contains no path between
$s_{i,j}$ and $t_{i,j}$.
\end{claim}

\begin{proof}
Fix $i\in\{1,2\}$, $j\in[\ell+1]$, and
$Z\subseteq\{x_b:b\in\mathcal B\}$.
Let $Q$ denote the added copy of
$K_{t+2}-s_{i,j}t_{i,j}$ on
$W_{i,j}\cup\{s_{i,j},t_{i,j}\}$.
Since $Z$ consists only of choice vertices, no vertex of
$Q$ belongs to $Z$.

Suppose first that $\widetilde H_i^{(j)}-Z$ contains an
$s_{i,j}$--$t_{i,j}$ path $P$.
The internal vertices of $P$ lie outside $V(Q)$.
Consequently, $Q\cup P$ is a subdivision of $K_{t+2}$:
the path $P$ replaces its missing terminal edge.
Thus $G_{i,j}-Z$ contains $K_{t+2}$ as a minor, and
$ \operatorname{tw}(G_{i,j}-Z)\ge t+1$.

Conversely, suppose that $\widetilde H_i^{(j)}-Z$ contains
no path between $s_{i,j}$ and $t_{i,j}$.
Since $\widetilde H_i^{(j)}$ is a subdivision of a
series-parallel graph, every component of
$\widetilde H_i^{(j)}-Z$ has treewidth at most two.
Moreover, each such component contains at most one terminal
and therefore intersects $Q$ in at most one vertex.
The graph $Q$ has a tree decomposition of width $t$
consisting of the two adjacent bags
$W_{i,j}\cup\{s_{i,j}\}$ and $W_{i,j}\cup\{t_{i,j}\}$.
For each component $C$ containing a terminal
$r\in\{s_{i,j},t_{i,j}\}$, take a tree decomposition of $C$
of width at most two. Choose a bag containing $r$ and
connect its node by an edge to the node with bag
$W_{i,j}\cup\{r\}$ in the decomposition of $Q$.
For each component containing neither terminal, take a tree
decomposition of width at most two and connect any node of
its decomposition tree to any node of the decomposition of $Q$.
This preserves the tree decomposition properties because
the component is disjoint from $Q$ and all other components.

Include each remaining isolated vertex of $G_{i,j}-Z$
in a singleton bag, attached to any node of the decomposition
tree. Every bag has size at most $\max\{t+1,3\}=t+1$,
since $t\ge2$. Hence
$\operatorname{tw}(G_{i,j}-Z)\le t$.
\end{proof}

\paragraph{Correctness.}
The forward direction is easy. 
Indeed, let $\mathcal B'\subseteq\mathcal B$ is a
solution to \textsc{Paired Cut} with $|\mathcal B'|\le \ell$.
Let
$Z:=\{x_b:b\in\mathcal B'\}$.
By construction, $\widetilde H_i^{(j)}-Z$ contains a path
between $s_{i,j}$ and $t_{i,j}$ if and only if
$H_i-\{e_i:(e_1,e_2)\in\mathcal B'\}$ contains a path
between $s_i$ and $t_i$.
Thus, for every $i\in\{1,2\}$ and $j\in[\ell+1]$,
$\widetilde H_i^{(j)}-Z$ contains no path between
$s_{i,j}$ and $t_{i,j}$.
From \cref{claim:network-treewidth} we conclude that
$\operatorname{tw}(G_{i,j}-Z)\le t$.
Since $|Z|=|\mathcal B'|\le\ell=K$, the set $Z$ is a solution for the instance $(V, \{G_{i,j}\}_{i \in [2], \\ j \in [\ell +1]}, K)$ of {\sc Simultaneous Treewidth-$t$ Deletion}.

For the backward direction,
let $D\subseteq V$ be a solution 
for the instance $(V, \{G_{i,j}\}_{i \in [2], j \in [\ell +1]}, K)$ of {\sc Simultaneous Treewidth-$t$ Deletion}, 
of size at most $K=\ell$.
Let $Z:=D\cap\{x_b:b\in\mathcal B\}$ and $\mathcal B':=\{b\in\mathcal B: x_b\in Z\}$.
Fix $i\in\{1,2\}$.
For each $j \in [\ell +1]$, the sets $V_{i,j}\cup W_{i,j}$,
are pairwise disjoint.
Since $|D|\le \ell$, there exists $j\in [\ell+1]$ such that
$D\cap(V_{i,j}\cup W_{i,j})=\emptyset$.
As $D\setminus Z$ contains no choice vertex, every vertex
of $D\setminus Z$ is therefore isolated in $G_{i,j}$.
Hence $G_{i,j}-Z$ differs from $G_{i,j}-D$
only in isolated vertices, and
$ \operatorname{tw}(G_{i,j}-Z)
  =\operatorname{tw}(G_{i,j}-D)\le t$.
By \cref{claim:network-treewidth},
$\widetilde H_i^{(j)}-Z$ contains no path between
$s_{i,j}$ and $t_{i,j}$.
By construction, deleting from $H_i$ the edges belonging
to pairs in $\mathcal B'$ therefore leaves no path between
$s_i$ and $t_i$.
For each $i\in\{1,2\}$, the graph
$H_i-\{e_i:(e_1,e_2)\in\mathcal B'\}$
contains no path between $s_i$ and $t_i$.
Moreover,
$|\mathcal{B'}|=|Z|\le|D|\le\ell$.
Thus $\mathcal B'$ is a solution to \textsc{Paired Cut}.

For fixed $t$, the construction has polynomial size and
takes polynomial time. Finally, the parameter of the new instance is 
$K+\alpha=\ell+2(\ell+1)=3\ell+2$
which
depends only on the parameter $\ell$ of the source problem.
This proves the lemma.
\end{proof}

\begin{proof}[Proof of \cref{thm:hardness}]
Fix $t\geq2$.
We give a parameterized reduction from
\textsc{Simultaneous Treewidth-$t$ Deletion}
to \strongbackt{}.
Let $(G_0,\ldots,G_{\alpha-1},K)$ be an instance of \textsc{Simultaneous Treewidth-$t$ Deletion},
where all graphs have common vertex set $V$.
By \cref{lem:simultaneous-treewidth-hardness},
\textsc{Simultaneous Treewidth-$t$ Deletion} is \Wone-hard parameterized
by $K+\alpha$.
We restrict to the instances produced by the reduction
in the proof of \cref{lem:simultaneous-treewidth-hardness}.
That reduction gives $K\geq1$ and $\alpha=2(K+1)$.
These instances also satisfy $K\leq|V|$.
Set $d:=\lceil\log\alpha\rceil$ and $k:=K+d$.
Since $\alpha\geq4$, we have $d\geq2$.

For every vertex $v\in V$, introduce a Boolean variable,
also denoted by $v$, and introduce selector
variables $S:=\{s_1,\ldots,s_d\}$.

  For each $a\in\{0,\ldots,\alpha-1\}$, let
$\operatorname{code}(a)\in\{0,1\}^d$ be the $d$-bit binary
representation of $a$. For every $i\in[d]$, define
$$
  \ell_{a,i}:=
  \begin{cases}
    s_i, & \text{if }\operatorname{code}(a)_i=0,\\
    \neg s_i, & \text{if }\operatorname{code}(a)_i=1.
  \end{cases}
$$
Thus all literals $\ell_{a,1},\ldots,\ell_{a,d}$ are false
under an assignment $\sigma:S\to\{0,1\}$ if and only if
$\sigma(s_i)=\operatorname{code}(a)_i$ for every $i\in[d]$.

\paragraph{Encoding the graph edges.}
For every $a\in\{0,\ldots,\alpha-1\}$ and edge $uv\in E(G_a)$,
choose an arbitrary orientation from $u$ to $v$ and
introduce fresh variables $z_1,\ldots,z_{d-1}$,
private to this edge occurrence.
For every $2 \leq i \leq d-1$, add the clause
$\{\neg z_{i-1},z_i,\ell_{a,i}\}$. Also add the clauses $\{u,z_1,\ell_{a,1}\}$ and $ \{\neg z_{d-1},v,\ell_{a,d}\}$.
We call these clauses the \emph{edge chain} for $uv$ in $G_a$.
The variables $z_1,\ldots,z_{d-1}$ are fresh for each
pair $(a,uv)$ with $a \in \{0,1, \ldots, \alpha-1\}$ and $uv\in E(G_a)$.
Let $\phi_0$ consist of all these clauses.

\paragraph{Forcing the selectors.}
For every $i\in[d]$ and $j\in[k+1]$, introduce fresh
variables $p_{i,j,1},\ldots,p_{i,j,t+1}$.
Add the clause $\{s_i,p_{i,j,r}\}$ for every
$r\in[t+1]$ and the clause
$\{p_{i,j,r},p_{i,j,r'}\}$ for every
$1\leq r<r'\leq t+1$.
These clauses correspond to all edges of a complete graph
on $\{s_i,p_{i,j,1},\ldots,p_{i,j,t+1}\}$.
Let $\phi$ be the union of $\phi_0$ and these clauses.

Observe that every clause is non-tautological and has size at most three.
Since $t$ is fixed, $d=\lceil\log\alpha\rceil$, and
$k=K+d$ with $K\le|V|$, the number of variables and clauses
is polynomial in the input size. The construction can
therefore be carried out in polynomial time.

\begin{claim}\label{claim:treewidth-selectors}
Every strong treewidth-$t$ backdoor of $\phi$ of size at most $k$ contains $S$.
\end{claim}

\begin{proof}
Let $X$ be such a backdoor and suppose that
$s_i\notin X$ for some $i\in[d]$.
For each $j\in[k+1]$, define
$P_j:=\{p_{i,j,1},\ldots,p_{i,j,t+1}\}$.
Since these sets are pairwise disjoint and $|X|\le k$,
there exists $j\in[k+1]$ such that $X\cap P_j=\emptyset$.
Thus $X\cap(\{s_i\}\cup P_j)=\emptyset$.

Fix an assignment $\beta:X\to\{0,1\}$.
All clauses corresponding to the edges of the complete
graph on $\{s_i\}\cup P_j$ survive unchanged in $\phi[\beta]$.
These clauses remain pairwise distinct, so
$\inc(\phi[\beta])$ contains a subdivision of $K_{t+2}$.
Consequently,
$ \operatorname{tw}(\inc(\phi[\beta]))\ge t+1$,
contradicting that $X$ is a strong treewidth-$t$ backdoor.
Hence $S\subseteq X$.
\end{proof}

\begin{claim}\label{claim:treewidth-edge-encoding}
Let $D\subseteq V\setminus S$, and let $\sigma:S\to\{0,1\}$ and
$\tau:D\to\{0,1\}$ be assignments.
If there exists $a\in\{0,\ldots,\alpha-1\}$ such that
$\sigma(s_i)=\operatorname{code}(a)_i$ for every $i\in[d]$,
then
$\operatorname{tw}\bigl(\inc(\phi_0[\sigma\cup\tau])\bigr)
  \le t$ if and only if 
  $\operatorname{tw}(G_a-D)\le t$.

If no such index $a$ exists, then
$\inc(\phi_0[\sigma\cup\tau])$ is a forest.
\end{claim}

\begin{proof}
Fix $D$, $\sigma$, and $\tau$ as in the statement of the claim.
Since $\sigma$ assigns only selector variables and $\tau$
assigns only variables corresponding to vertices in $D$,
all chain variables remain unassigned.
Every clause of $\phi_0[\sigma]$ or
$\phi_0[\sigma\cup\tau]$ therefore contains a literal on
a chain variable. Each chain variable occurs positively
in exactly one clause of $\phi_0$ and negatively in exactly
one other clause. Thus each literal on a chain variable
occurs in exactly one clause of $\phi_0$.
Consequently, distinct clauses of $\phi_0$ that are not
deleted yield distinct clauses in each of these reducts.

Suppose first that there exists
$a\in\{0,\ldots,\alpha-1\}$ such that
$\sigma(s_i)=\operatorname{code}(a)_i$ for every $i\in[d]$.
This index is unique because the binary strings
$\operatorname{code}(0),\ldots,\operatorname{code}(\alpha-1)$
are pairwise distinct.
By definition, every literal $\ell_{a,i}$ is false under
$\sigma$. Hence the clauses introduced for each edge of
$G_a$ become binary in $\phi_0[\sigma]$.

For an edge $uv\in E(G_a)$, let $C_1,\ldots,C_d$ denote
the vertices corresponding to these binary clauses.
The vertices $C_1,\ldots,C_d$ and the vertices labelled
$u,z_1,\ldots,z_{d-1},v$ form the path
$ u,C_1,z_1,C_2,\ldots, \allowbreak z_{d-1},C_d,v $
in $\inc(\phi_0[\sigma])$.
If $u,v\notin D$, none of the variables
$u,z_1,\ldots,z_{d-1},v$ is assigned by $\tau$.
Thus this path is also present in
$\inc(\phi_0[\sigma\cup\tau])$.
If $u\in D$, the clause $\{u,z_1\}$ is deleted in
$\phi_0[\sigma\cup\tau]$ when $\tau(u)=1$ and becomes
the unit clause $\{z_1\}$ when $\tau(u)=0$.
In either case, the vertex labelled $u$ is absent from
$\inc(\phi_0[\sigma\cup\tau])$.
The same argument applies to $v$ and the clause
$\{\neg z_{d-1},v\}$.
Consequently, every edge of $G_a-D$ gives a path between
its endpoints in $\inc(\phi_0[\sigma\cup\tau])$.
For every other edge of $G_a$, the remaining vertices and
edges corresponding to its chain form paths, each containing
at most one vertex labelled by an element of $V\setminus D$.

For every $b\in\{0,\ldots,\alpha-1\}\setminus\{a\}$,
there exists $i\in[d]$ such that
$\sigma(s_i)\ne\operatorname{code}(b)_i$.
Hence $\ell_{b,i}$ is true under $\sigma$.
For each edge $uv\in E(G_b)$, the clause containing
$\ell_{b,i}$ is therefore satisfied by $\sigma$ and deleted
in $\phi_0[\sigma\cup\tau]$.
The vertex corresponding to that clause is absent from
$\inc(\phi_0[\sigma\cup\tau])$.
The remaining clause vertices of this chain, their neighbors,
and their incident edges therefore form disjoint paths,
each containing at most one of the endpoint vertices $u,v$.

The paths corresponding to edges of $G_a-D$ have pairwise
disjoint internal vertices.
Every other remaining path contains at most one vertex
labelled by an element of $V\setminus D$, and all its other
vertices occur in no other path.
Thus each connected component formed by these other paths
is a tree containing at most one vertex labelled by an
element of $V\setminus D$.
It can intersect the paths corresponding to $G_a-D$ only
at that vertex.
Replacing edges by paths and attaching trees at single
vertices preserve whether treewidth is at most $t$,
since $t\ge2$. Hence
$\operatorname{tw}\bigl(\inc(\phi_0[\sigma\cup\tau])\bigr) \le t
$
 if and only if 
$\operatorname{tw}(G_a-D)\le t$.

Finally, suppose that no index $a\in\{0,\ldots,\alpha-1\}$
satisfies $\sigma(s_i)=\operatorname{code}(a)_i$ for every
$i\in[d]$.
Then, for each $b\in\{0,\ldots,\alpha-1\}$, some literal
$\ell_{b,i}$ is true under $\sigma$.
Every edge chain therefore has a clause deleted in
$\phi_0[\sigma\cup\tau]$.
As above, the remaining clause vertices, their neighbors,
and their incident edges form paths, each containing at
most one vertex labelled by an element of $V\setminus D$.
By the preceding argument, these paths form tree components.
Hence $\inc(\phi_0[\sigma\cup\tau])$ is a forest.
\end{proof}

\paragraph{Correctness.}
For the forward direction, let $D\subseteq V$ satisfy
$|D|\le K$ and $\operatorname{tw}(G_a-D)\le t$ for every
$a\in\{0,\ldots,\alpha-1\}$.
We may assume that $D$ contains no vertex isolated in every
input graph, since removing such vertices from $D$ preserves
$\operatorname{tw}(G_a-D)\le t$ for every
$a\in\{0,\ldots,\alpha-1\}$.
Set $X:=S\cup D$ (recall $S$ is the set of selector variables). Then $|X|\le d+K=k$.

Fix an arbitrary assignment $\beta:X\to\{0,1\}$, and put
$\sigma:=\beta|_S$ and $\tau:=\beta|_D$.
If there exists $a\in\{0,\ldots,\alpha-1\}$ such that
$\sigma(s_i)=\operatorname{code}(a)_i$ for every $i\in[d]$,
then \cref{claim:treewidth-edge-encoding} gives
$\operatorname{tw}\bigl(\inc(\phi_0[\beta])\bigr)\le t$,
since $\operatorname{tw}(G_a-D)\le t$.
If no such index $a$ exists, the same claim shows that
$\inc(\phi_0[\beta])$ is a forest and therefore has
treewidth at most one. Since $t\ge2$, we again obtain
$\operatorname{tw}(\inc(\phi_0[\beta]))\le t$.

Consider the clauses introduced for a fixed $i\in[d]$ and
$j\in[k+1]$ in the selector-forcing construction.
Their variables other than $s_i$ remain unassigned by $\beta$.
Among these clauses, those containing $s_i$ are deleted
when $\beta(s_i)=1$ and become unit clauses when $\beta(s_i)=0$.
The clauses on pairs of the other $t+1$ variables remain
unchanged. In $\inc(\phi[\beta])$, the corresponding clause
vertices and their neighbors therefore form a subdivision
of $K_{t+1}$, with possible leaves corresponding to unit
clauses. This graph has treewidth $t$.
The sets $\{p_{i,j,1},\ldots,p_{i,j,t+1}\}$ are pairwise
disjoint over all $(i,j)\in[d]\times[k+1]$.
None of these variables occurs in $\phi_0$.
Consequently, each such graph is a separate component of
$\inc(\phi[\beta])$.
Since $\operatorname{tw}(\inc(\phi_0[\beta]))\le t$
and every additional component has treewidth at most $t$,
we obtain
$\operatorname{tw}(\inc(\phi[\beta]))\le t$.
Since $\beta$ was arbitrary, $X$ is a strong treewidth-$t$ backdoor of
size at most $k$.

For the backward direction, let $X$ be a strong
treewidth-$t$ backdoor of $\phi$ of size at most $k$.
By \cref{claim:treewidth-selectors}, $S\subseteq X$.
Let $Y:=X\setminus S$, therefore $|Y|\le K$.

Construct $D\subseteq V$ as follows.
For every $v\in V$, include $v$ in $D$ if its corresponding
variable belongs to $Y$.
For every chain variable in $Y$, choose one endpoint of
the edge for which that variable was introduced and include
that endpoint in $D$.
Variables introduced in the selector-forcing construction
contribute no vertices to $D$.
Each variable of $Y$ contributes at most one vertex, so
$|D|\le|Y|\le K$.

Fix $a\in\{0,\ldots,\alpha-1\}$ and choose an assignment
$\beta:X\to\{0,1\}$ satisfying
$\beta(s_i)=\operatorname{code}(a)_i$ for every $i\in[d]$.
Consider an edge $uv\in E(G_a-D)$.
Neither of the variables $u,v$ belongs to $Y$, by the
construction of $D$.
No chain variable introduced for this edge belongs to $Y$
either, since otherwise an endpoint of $uv$ would belong
to $D$.
Thus the clauses introduced for this edge become binary
in $\phi[\beta]$, and their corresponding clause vertices
and neighboring vertices form a path from $u$ to $v$ in
$\inc(\phi[\beta])$.

Each of these binary clauses contains a literal $\ell$
on an unassigned chain variable, where $\ell$ occurs
in exactly one clause of $\phi$.
Consequently, these clauses remain distinct in $\phi[\beta]$.
Since different edge chains use distinct chain variables,
the resulting paths have pairwise disjoint internal vertices.
Each connected component of $G_a-D$ containing an edge is
therefore a minor of $\inc(\phi[\beta])$.
Since $X$ is a strong backdoor,
$\operatorname{tw}(\inc(\phi[\beta]))\le t$, and hence
$\operatorname{tw}(G_a-D)\le t$.
As $a$ was arbitrary, $D$ is a simultaneous deletion set
of size at most $K$. This proves the equivalence.

\paragraph{Parameter and running time.}
Observe that the construction takes polynomial time and produces a
non-tautological $3$-CNF formula. Its backdoor budget satisfies
$ k=K+\lceil\log\alpha\rceil\le K+\alpha$.
Thus the target parameter is bounded by the source parameter
$K+\alpha$. The reduction therefore establishes \Wone-hardness
of \strongbackt{} parameterized by $k$.

\end{proof}

\section{Conclusion}\label{sec:conclusion}

We determine the parameterized complexity of finding a set of at
most $k$ variables of a non-tautological CNF formula such that
every assignment to these variables leaves a formula whose
incidence graph has treewidth at most $t$.
When $t=1$, so that every residual incidence graph must be a
forest, we give an algorithm running in
$2^{2^{O(k)}}\cdot|\phi|$ time.
For every fixed $t\geq 2$, finding such a set is \Wone-hard
parameterized by $k$, even on non-tautological $3$-CNF formulas.
Thus, allowing the residual incidence graphs to have treewidth
two instead of one changes the detection problem from FPT to
\Wone-hard.

For the target class of incidence forests, we also relate the
dependence on backdoor size in the running time to the complexity of satisfiability.
An algorithm finding a strong backdoor of size at most $k$ in
$2^{2^{o(k)}}\cdot|\phi|^{O(1)}$ time would yield an algorithm
solving $3$-SAT on $N$ variables in $2^{o(N/\log N)}$ time, which would mean an even faster algorithm for solving {\sc $3$-SAT} than what is currently claimed by OpenAI~\cite{OpenAI2026ETH} and Anthropic~\cite{Claude2026ETH}, in the result where they claim to refute ETH.
This way we state the consequence of improving our parameter dependence
directly, without relying on ETH.

For every fixed target treewidth $t$, the FPT-approximation
algorithm of Gaspers and Szeider~\cite{DBLP:conf/focs/GaspersS13}
finds a strong backdoor of size at most $2^k$ whenever one of size
at most $k$ exists. Such a backdoor suffices to solve SAT in FPT
time, even though its size may be exponentially larger than the
requested bound. 
For these larger target treewidth, 
a natural next question is
whether better FPT-approximations are possible. 
In particular,
can an FPT in $k+t$ algorithm find a strong backdoor of size at most
$k^{O(1)}$ or even $k^{O(\log k)}$ whenever a strong backdoor of size at most $k$ exists,
on arbitrary non-tautological CNF formulas, when the target class has treewidth $t$?

\paragraph*{AI disclosure.}
AI was used only for editing and polishing the proofs after having the preliminary proofs written by the authors.
The authors take full responsibility for the content of this work.

\clearpage
\bibliographystyle{alpha}
\bibliography{references}

\appendix

\section{Allowing tautological clauses in $3$-CNF formulas}
\label{sec:tautological-3cnf}

We consider \textsc{Simultaneous Colored Strong Backdoor
to Forests}, now allowing tautological clauses.
Recall an instance of this problem
 is $I=(\Phi,V,P_1,\ldots,P_k)$, where
$\Phi=(\phi_1,\ldots,\phi_\alpha)$ is a collection of $\alpha$ $3$-CNF
formulas over $V$.
The goal is to find $x_j\in P_j$, with repetitions allowed,
such that $X=\{x_1,\ldots,x_k\}$ is a strong forest
backdoor of every formula in $\Phi$.

\begin{theorem}
Let $I=(\Phi,V,P_1,\ldots,P_k)$ be an instance of
\textsc{Simultaneous Colored Strong Backdoor to Forests},
where $\Phi$ contains $\alpha$ $3$-CNF formulas,
possibly with tautological clauses.
Then $I$ can be solved in
$2^{\alpha2^{O(k)}}\cdot|I|$ time.
\end{theorem}

\begin{proof}
We first show how to replace tautological clauses
by non-tautological clauses once the incidence
graphs are $K_{2,2}$-free.
We then give a bounded branching procedure for
reaching this case.

\paragraph{The $K_{2,2}$-free case.}
Let $\psi$ be a $3$-CNF formula whose incidence
graph is $K_{2,2}$-free.
Every tautological clause contains at most two
distinct variables.
Obtain $\widehat{\psi}$ by deleting one of the
two complementary literals from each tautological
clause and merging identical clauses.
The formula $\widehat{\psi}$ is non-tautological,
and each modified clause retains its original
variable set.
Only unit clauses can newly coincide: otherwise,
two original clauses would share two distinct
variables, contradicting $K_{2,2}$-freeness.
Thus $\inc(\widehat{\psi})$ is also $K_{2,2}$-free.
This transformation takes linear time.

We claim that every $X\subseteq V$ is a strong
forest backdoor of $\psi$ if and only if it is
one of $\widehat{\psi}$.
Fix an assignment $\tau$ to $X$ and consider an
original tautological clause $C$.
If $X\cap\var(C)=\emptyset$, both $C$ and its
modified clause survive with the same variable
neighbors.
Otherwise, at most one variable of $C$ remains
unassigned.
In this case, the corresponding clause vertex
in either semi-reduct is absent or has degree
at most one.
All non-tautological clauses are treated identically.
Consequently, the two semi-reduct incidence graphs
are acyclic simultaneously.

Moreover, in either formula, clauses that become
identical after an assignment can contain at most
one distinct variable, by $K_{2,2}$-freeness.
Merging their clause vertices therefore preserves
acyclicity.
Hence
\[
    \inc(\psi[\tau]) \text{ is a forest}
    \quad\Longleftrightarrow\quad
    \inc(\widehat{\psi}[\tau]) \text{ is a forest}.
\]
This proves the claim.

Apply this transformation separately to every
input formula, keeping $V$ and all candidate
sets unchanged.
The validity of every choice from the candidate
sets is preserved.
We may therefore apply \cref{lem:colorful-linear}
directly to the resulting non-tautological instance.

\paragraph{Eliminating four-cycles.}
At each recursive call, reject if a candidate
set $P_j$
is empty.
If no candidate sets remain, accept precisely
when every incidence graph is a forest.
If all incidence graphs are $K_{2,2}$-free, use
the procedure established above.

Otherwise, find a four-cycle, using Lemma~\ref{lem:find-k22}, through two distinct
clauses $C_1,C_2$.
Every solution intersects
$\var(C_1)\cup\var(C_2)$.
Indeed, a choice avoiding these variables leaves
both clauses unchanged and distinct under every
assignment, so the four-cycle survives.
Since each clause contains at most three literals,
this union has at most six variables.

For every $v\in\var(C_1)\cup\var(C_2)$ and every
remaining candidate set $P_j$ containing $v$,
create a branch choosing $v$ from $P_j$.
Remove $P_j$ and replace every current formula
$\psi$ by both $\psi[v=0]$ and $\psi[v=1]$.
All other
candidate sets unchanged.
The remaining choices must work for both
assignments to $v$, so this branching is sound
and complete.
This also permits $v$ to be chosen from another
candidate set, since it is absent from both reducts.

Each branch removes one candidate set.
The search tree therefore has depth at most $k$,
at most $6k$ children per node, and
$2^{O(k\log(k+2))}$ nodes.
At depth $r$, there are at most $\alpha2^r$
formulas, at most $k-r$ candidate sets, and
the instance size is $2^{O(r)}|I|$.
Every incidence graph is $3$-degenerate, so
\cref{lem:find-k22} finds a four-cycle, if one
exists, in linear time.
Constructing the reducts also takes linear time
in the current instance size.

At a $K_{2,2}$-free leaf, first replace each
tautological clause by a non-tautological clause
as described above, and then apply
\cref{lem:colorful-linear}.
The running time is at most
\[
    2^{\,\alpha2^r2^{O(k-r)}}\,
    2^{O(r)}|I|
    \leq
    2^{\alpha2^{O(k)}}|I|.
\]
Summing over the search tree gives the claimed
running time.
Successful choices are recovered from the leaf
algorithm and the variables fixed along its branch.
\end{proof}

For a single input formula $\phi$, first handle
$k=0$ or an already acyclic incidence graph
directly.
Otherwise, set $\alpha=1$, $V=\var(\phi)$,
and $P_j=V$ for every $j\in[k]$.
As in the proof of \cref{thm:main}, repeated
choices represent every nonempty backdoor of
size at most $k$.
The resulting running time is
$2^{2^{O(k)}}|\phi|$.
Thus \cref{thm:main} also holds for $3$-CNF
formulas that may contain tautological clauses.

\end{document}